\documentclass[11pt]{article}
\usepackage{amsmath}
\usepackage{amsfonts}
\usepackage{amssymb}
\usepackage{graphicx}
\usepackage{color}
\usepackage[table,xcdraw]{xcolor}
\usepackage{orcidlink}
\usepackage{hyperref}
\hypersetup{colorlinks=true, linkcolor=blue, citecolor=blue, urlcolor=blue}
\usepackage{xcolor}
\usepackage{xstring}
\newcommand{\hrefbibentry}[2]{%
  \IfStrEq{#1}{}{#2}{%
    \IfBeginWith{#1}{http}%
      {\href{#1}{\textcolor{blue}{#2}}}%
      {\href{https://doi.org/#1}{\textcolor{blue}{#2}}}}}
\usepackage{float}
\usepackage{multirow}
\usepackage{array}
\usepackage{longtable}
\usepackage{booktabs}
\usepackage{colortbl}
\usepackage{tabularx}
\usepackage{xltabular}
\newcolumntype{C}{>{\centering\arraybackslash}X}
\newcolumntype{L}{>{\raggedright\arraybackslash}X}
\newcolumntype{R}{>{\raggedleft\arraybackslash}X}
\usepackage{geometry}
\usepackage[font=small,labelfont=bf,skip=4pt]{caption}
\usepackage{microtype}
\RequirePackage[numbers,sort&compress]{natbib}
\numberwithin{equation}{section}
\graphicspath{{Figures/}}
\newcommand{\dd}{\mathrm{d}}
\newcommand{\rH}{r_{\rm H}}

\newcommand{\Ms}{M}
\newcommand{\wH}{w_{\rm H}}

\begin{document}
\baselineskip=20pt
\begin{center}
\LARGE{An exact dyonic black hole at the integrable locus of quadratic nonlinear electrodynamics}
\end{center}
\vspace{0.2cm}

\begin{center}
{\bf Faizuddin Ahmed\orcidlink{0000-0003-2196-9622}}\footnote{\bf fahmed@rgu.ac ; faizuddinahmed15@gmail.com}$^{a}$,\ \
{\bf Ahmad Al-Badawi\orcidlink{0000-0002-3127-3453}}\footnote{\bf ahmadbadawi@ahu.edu.jo}$^{b}$\ \ and\ \
{\bf \.{I}zzet Sakall{\i}\orcidlink{0000-0001-7827-9476}}\footnote{\bf izzet.sakalli@emu.edu.tr (Corresponding author)}$^{c}$\\[3pt]
{\it $^{a}$Department of Physics, The Assam Royal Global University, Guwahati, 781035, Assam, India}\\
{\it $^{b}$Department of Physics, Al-Hussein Bin Talal University, 71111, Ma'an, Jordan}\\
{\it $^{c}$Physics Department, Eastern Mediterranean University, Famagusta 99628, North Cyprus via Mersin 10, Turkey}
\end{center}
\vspace{0.2cm}

\begin{abstract}
Black holes carrying both electric and magnetic charge in the quadratic nonlinear electrodynamics defined by $-F+aF^{2}+bG^{2}$ have so far been obtained in closed form only when the $F^{2}$ coupling is switched off, and are otherwise treated perturbatively or numerically, because the electric field solves a cubic whose coefficients depend on the radius. We identify the locus $b=a/2$, on which that radial dependence cancels, the magnetic charge decouples from the constitutive relation, and the field equations integrate exactly. Using the field strength rather than the areal radius as the independent variable, we obtain a dyonic black hole in parametric closed form, with the radial quadrature evaluated as a sum of Gauss hypergeometric functions and with the cosmological constant carried throughout. The solution reduces to dyonic Reissner--Nordstr\"om--(anti-)de Sitter in the Maxwell limit, to the known magnetic solution when the electric charge vanishes, and to the Euler--Heisenberg electric branch when the magnetic charge vanishes; at first order in the couplings it reproduces the perturbative dyonic lapse function, which it extends to all orders, and it is invariant under the interchange of the two charges to that order, the asymmetry entering only at second order. Its horizon structure contains a three-horizon window that requires both charges and closes off in either of them. The same change of variable renders the electric Smarr integral, previously stated to have no closed form, as a combination of hypergeometric functions. We obtain the energy density and the pressures in closed form, from which the electric charge is seen to enlarge the region where the dominant energy condition holds beyond the purely magnetic bound $\rH^{4}>4ap^{2}$, and we identify three distinct curvature laws at the centre, one of which is a relegated singularity produced by tuning the mass to the electric self-energy. The extended thermodynamics closes with the coupling and the pressure as variables, the conjugate of the coupling being a new hypergeometric quadrature, and the critical ratio falls below the Maxwell value $3/8$ and disappears altogether above a threshold coupling. Finally, the two photon cones at the locus separate the shadow into a doublet whose splitting reaches roughly half a percent, while the null geodesic light ring and the scalar ringdown remain almost blind to the coupling, so the nonlinearity is carried by the polarization structure rather than by the geometry that a shadow diameter measures.
\end{abstract}


\section{Introduction}\label{isec1}

Nonlinear electrodynamics entered gravity as a way of taming the self-energy of a point charge \citep{Born1934}, and it has since supplied most of the known regular black hole geometries \citep{AyonBeato1998,Bronnikov2001,Fan2016,Bronnikov2018}. Of the several ways of organizing the nonlinearity, the quadratic Lagrangian in the two invariants $F$ and $G$ occupies a useful middle ground. It contains the one-loop Euler--Heisenberg theory \citep{HeisenbergEuler1936} and the leading Born--Infeld correction as particular choices of the two couplings, and it is simple enough that the field equations sometimes close without further approximation.

Sometimes, but rarely, and the pattern of what closes is instructive. When the electric charge vanishes the constitutive relation is trivial and the magnetic black hole is elementary \citep{Croney2025}. When the $F^{2}$ coupling vanishes the cubic that fixes the electric field degenerates to a linear relation and the dyonic solution closes in logarithms and arctangents \citep{Croney2026,Wang2025}. For a purely electric configuration the cubic can be solved by Cardano's formula and the quadrature evaluated in hypergeometric functions, which was carried out recently for the Euler--Heisenberg couplings \citep{Luo2026}, after earlier treatments had either worked to first order in the coupling \citep{Yajima2001} or passed to the dual formulation in which the auxiliary invariant, rather than the physical one, is the independent variable \citep{Magos2020,Breton2021}. What has resisted all of these is the dyonic configuration at non-vanishing $F^{2}$ coupling, where the cubic carries an $r$-dependent coefficient sourced by the magnetic charge. Both of the recent detailed studies of this sector report it as analytically intractable and proceed numerically \citep{Croney2026,Luo2026}, and what is otherwise available is the lapse function to first order in the couplings, which is accurate far from the horizon and degrades exactly where the horizon structure is interesting.

The dyonic sector is not a curiosity. Two charges rather than one change the thermodynamics qualitatively, producing triple points and reentrant behaviour that the single-charge theory does not have \citep{Li2022a,Mou2024,Chen2024}, and they change the optics as well, since a background with both an electric and a magnetic field has a non-vanishing pseudoscalar invariant and is therefore birefringent \citep{Breton2021,Paula2026}. Exact dyonic solutions are consequently in demand across a range of nonlinear theories, from Born--Infeld and its generalisations \citep{Yang2022,Yang2023,Li2016,Kruglov2019a,Acuna2026} to logarithmic models \citep{Kruglov2019}, conformally invariant electrodynamics \citep{Kokoska2021,Ortaggio2026}, quasitopological electromagnetism \citep{Ali2024,Sekhmani2023}, Kaluza--Klein constructions \citep{Mignemi2022}, noncommutative geometry \citep{Bokuli2026}, ModMax with a Kalb--Ramond background \citep{Ahmed2026,Ahmed2026a,ALBADAWI2025102076}, and general classifications of when a dyonic configuration can be nonsingular at all \citep{Tsuda2026,Bronnikov2017}. Within the quadratic family itself, however, the dyonic problem at $a\neq0$ has stayed open.

The obstruction is narrower than it looks. The coefficient in question is proportional to $2b-a$, so it vanishes identically on a one-dimensional locus in the space of couplings, and on that locus the magnetic charge drops out of the constitutive relation while remaining in the geometry. The electric field then solves a cubic with constant coefficients, which can be inverted to give the radius algebraically, and the quadrature that fixes the metric function becomes an ordinary integral of an algebraic function. This is the construction we carry out here. It produces an exact dyonic black hole with both couplings nonzero, the first of its kind in this family, and it does so at the cost of one relation between the couplings rather than at the cost of a perturbative expansion. Since the cosmological constant does not enter the gauge field equation, the construction carries it without modification, and we keep it throughout so that the extended thermodynamics is available.

The locus $b=a/2$ is not the Euler--Heisenberg ratio $b=7a/4$ nor the Born--Infeld value $b=a$, and we do not claim it is selected by any deeper principle. What recommends it is that it is the only place in the quadratic family where the dyonic problem is exactly soluble, so it provides the controlled reference point that the numerical and perturbative treatments elsewhere in this sector currently lack, in the same way that exactly soluble points are used throughout the black hole literature. One structural feature does emerge on its own. Through first order in the coupling the lapse function at the locus is symmetric under the interchange of the electric and magnetic charges, and we verify in Sec.~\ref{isec3} that the asymmetry appears only at second order, with a coefficient we compute.

Section~\ref{isec2} fixes the theory, derives the integrability condition and introduces the change of variable it permits. Section~\ref{isec3} presents the dyonic solution, checks it against every limit in which the answer is already known, places it against the perturbative lapse function used elsewhere in this sector, and then examines the horizon structure and the core in turn. Section~\ref{isec4} recovers the electric branch, closes the Smarr integral analytically, and develops the thermodynamics with the coupling and the pressure promoted to thermodynamic variables. Section~\ref{isec5} derives the photon cones and the birefringence they produce, and Sec.~\ref{isec6} the ringdown, the greybody factors and the shadow. Section~\ref{isec7} concludes. Appendix~\ref{app:A} collects the closed forms of the radial integrals and Appendix~\ref{app:B} the numerical verification. Geometrised units $G=c=\hbar=k_{B}=1$ and the signature $(-,+,+,+)$ are used throughout.

\section{Quadratic nonlinear electrodynamics with two charges and its integrable locus}\label{isec2}

The theory and the dyonic field equations are collected here in the form used throughout, following Refs.~\citep{Croney2025,Croney2026}, so that the departure taken in Sec.~\ref{isec2} can be stated against a definite starting point. Nothing in this section is new; what matters is which term carries the obstruction.

The action is
\begin{equation}
S=\frac{1}{16\pi}\int \dd^{4}x\,\sqrt{-g}\,\left(R-2\Lambda-F+aF^{2}+bG^{2}\right),
\qquad F\equiv F_{\mu\nu}F^{\mu\nu},\quad G\equiv F_{\mu\nu}\tilde{F}^{\mu\nu},
\label{eq:action}
\end{equation}
and the metric is taken in the static form with $g_{tt}g_{rr}=-1$,
\begin{equation}
\dd s^{2}=-f(r)\,\dd t^{2}+\frac{\dd r^{2}}{f(r)}+r^{2}\left(\dd\theta^{2}+\sin^{2}\theta\,\dd\varphi^{2}\right).
\label{eq:metric}
\end{equation}
Setting $a=b=\Lambda=0$ returns Einstein--Maxwell theory, while the Euler--Heisenberg and truncated Born--Infeld theories correspond to $b=7a/4$ and $b=a$ respectively. The Lagrangian derivatives that will be needed later are
\begin{equation}
\mathcal{L}_{F}=-1+2aF,\qquad \mathcal{L}_{FF}=2a,\qquad \mathcal{L}_{G}=2bG,\qquad
\mathcal{L}_{GG}=2b,\qquad \mathcal{L}_{FG}=0 ,
\label{eq:Lderivs}
\end{equation}
the vanishing of the mixed derivative being a property of the quadratic truncation and not of the locus.

A dyonic configuration carries the magnetic field $F_{\theta\varphi}=p\sin\theta$ together with a radial electric field $F_{rt}$. The cosmological constant does not appear in the gauge field equation, so $\nabla_{\mu}P^{\mu\nu}=0$ integrates to $P^{rt}=q/r^{2}$ as before, which defines the electric charge measured at infinity, and eliminating $P^{rt}$ leaves a cubic for the field strength,
\begin{equation}
4a\,F_{rt}^{3}+\left[1+\frac{4p^{2}}{r^{4}}\left(2b-a\right)\right]F_{rt}-\frac{q}{r^{2}}=0 .
\label{eq:cubic}
\end{equation}
For $a>0$ the cubic is monotone in $F_{rt}$, so its real root is unique and joins continuously to the Maxwell value $q/r^{2}$. The Einstein equations then integrate to
\begin{equation}
f(r)=1-\frac{2m}{r}-\frac{\Lambda r^{2}}{3}+\frac{p^{2}}{r^{2}}-\frac{2ap^{4}}{5r^{6}}
+\frac{1}{r}\int_{r}^{\infty}\dd r'\left[2a\,r'^{2}F_{rt}^{4}+q\,F_{rt}\right],
\label{eq:master}
\end{equation}
the cosmological constant entering only through the term $-\Lambda r^{2}/3$.

Two features of \eqref{eq:master} drive what follows. The integrand contains no explicit magnetic charge, so $p$ enters the quadrature only through the bracket in \eqref{eq:cubic}; and the magnetic contribution to $f$ is already elementary, so the entire difficulty of the dyonic problem sits in that bracket. Setting $q=0$ kills the integral and returns the magnetic solution of Ref.~\citep{Croney2025} with a cosmological constant,
\begin{equation}
f(r)=1-\frac{2m}{r}-\frac{\Lambda r^{2}}{3}+\frac{p^{2}}{r^{2}}-\frac{2ap^{4}}{5r^{6}},
\label{eq:magnetic}
\end{equation}
which is independent of $b$ because $G$ vanishes on a purely magnetic background. Setting $a=0$ instead reduces \eqref{eq:cubic} to $F_{rt}=qr^{2}/(r^{4}+8bp^{2})$ and the quadrature to elementary functions, giving the quasitopological dyonic solution \citep{Croney2026,Wang2025}. Everything else in this sector has been treated perturbatively or numerically.

The radial dependence of the bracket in \eqref{eq:cubic} is what prevents $F_{rt}$ from being inverted usefully. We now identify where that dependence disappears and show that the system then closes. The bracket equals unity in two circumstances,
\begin{equation}
p=0 \quad\text{(any $a$, $b$)},\qquad\text{or}\qquad b=\frac{a}{2}\quad\text{(any $p$, $q$)}.
\label{eq:loci}
\end{equation}
The first is a statement about the configuration and reproduces the purely electric problem, in which $b$ is absent altogether since $G=0$; this is the case solved in Ref.~\citep{Luo2026} and recovered in Sec.~\ref{isec4}. The second is a statement about the theory. On it the magnetic charge decouples from the constitutive relation, so the electric field obeys the same cubic it would obey with no magnetic monopole present, while the magnetic charge continues to gravitate through the explicit terms of \eqref{eq:master}. That separation is what makes the dyonic problem soluble, and it is available nowhere else in the quadratic family.

On either locus \eqref{eq:cubic} reads $4aF_{rt}^{3}+F_{rt}=q/r^{2}$. Writing $F_{rt}=w^{2}$ with $w>0$ and solving for the radius gives the algebraic relation
\begin{equation}
r(w)=\frac{\sqrt{q}}{w\sqrt{1+4aw^{4}}},
\qquad
\frac{\dd r}{\dd w}=-\frac{\sqrt{q}\,(1+12aw^{4})}{w^{2}(1+4aw^{4})^{3/2}} ,
\label{eq:rofw}
\end{equation}
which is strictly decreasing for $a>0$, so $w\in(0,\infty)$ covers $r\in(\infty,0)$ once and the parametrization is global. The Maxwell limit corresponds to $w=\sqrt{q}/r$. Since \eqref{eq:rofw} is monotone the parametrization carries the same information as an explicit $f(r)$, and every geometric quantity built from $f$ and its derivatives follows by the chain rule; the explicit first and second derivatives are recorded in Appendix~\ref{app:A}. The relation \eqref{eq:rofw} involves neither $p$ nor $\Lambda$, so the same change of variable serves the dyonic problem and its (anti-)de Sitter completion.

Changing the integration variable in \eqref{eq:master} by means of \eqref{eq:rofw}, the integrand becomes an algebraic function of $w$,
\begin{equation}
\left[2a\,r'^{2}F_{rt}^{4}+q\,F_{rt}\right]\dd r'
=-\,q^{3/2}\,\frac{(1+6aw^{4})(1+12aw^{4})}{(1+4aw^{4})^{5/2}}\,\dd w ,
\label{eq:integrand}
\end{equation}
so that the quadrature reduces to
\begin{equation}
\frac{1}{r}\int_{r}^{\infty}\dd r'\left[2a\,r'^{2}F_{rt}^{4}+q\,F_{rt}\right]=\frac{q^{3/2}}{r(w)}\,J(w),
\qquad
J(w)=\int_{0}^{w}\frac{72a^{2}s^{8}+18as^{4}+1}{(1+4as^{4})^{5/2}}\,\dd s .
\label{eq:Jdef}
\end{equation}
The closed form of $J$ in Gauss hypergeometric functions, and its limiting value $J_{\infty}$ in Beta functions, are given in Appendix~\ref{app:A}. The device of trading the areal radius for a field variable is what makes the whole of the rest of this paper elementary rather than numerical, and it is worth saying why it works here and not in general. The cubic at the locus has constant coefficients, so $r$ is an algebraic function of $F_{rt}$; off the locus the coefficients carry $r$ itself, the relation is no longer algebraic, and no such inversion exists.

\section{The exact dyonic black hole, its limits and its horizons}\label{isec3}

Combining \eqref{eq:master} with \eqref{eq:Jdef} at $b=a/2$ gives the solution that is the subject of this paper. We record it, verify that it degenerates correctly in every corner of parameter space where the answer is independently known, and then place it against the perturbative lapse function that is otherwise used in this sector.

The metric function is
\begin{equation}
f=1-\frac{2m}{r}-\frac{\Lambda r^{2}}{3}+\frac{p^{2}}{r^{2}}-\frac{2ap^{4}}{5r^{6}}+\frac{q^{3/2}J(w)}{r},
\qquad r=r(w),
\label{eq:dyonic}
\end{equation}
with $r(w)$ and $J(w)$ from \eqref{eq:rofw} and \eqref{eq:Jdef}. Four limits check it immediately. For $a\to0$ at fixed $b=a/2$ the integrand of \eqref{eq:Jdef} tends to unity, so $J(w)\to w$ while \eqref{eq:rofw} gives $w\to\sqrt{q}/r$, and \eqref{eq:dyonic} becomes
\begin{equation}
f\to 1-\frac{2m}{r}-\frac{\Lambda r^{2}}{3}+\frac{p^{2}+q^{2}}{r^{2}},
\label{eq:RNlimit}
\end{equation}
the dyonic Reissner--Nordstr\"om--(anti-)de Sitter lapse function. Setting $q\to0$ at fixed $r$ sends $w\to0$ with $q^{3/2}J/r\to q^{2}/r^{2}\to0$ and returns the magnetic solution \eqref{eq:magnetic}. Setting $p\to0$ returns the electric solution of Sec.~\ref{isec4}. Setting both charges to zero leaves Schwarzschild--(anti-)de Sitter. Table~\ref{tab:limits} reports these checks numerically together with the remaining ones, including the two series expansions of Appendix~\ref{app:A}, the agreement of the parametric form with direct numerical treatment of \eqref{eq:cubic} and \eqref{eq:master}, and a control run off the locus in which the two must and do disagree.

\begin{table}[!htbp]
\centering
\caption{Every limit in which the answer is independently known, checked against the parametric solution \eqref{eq:dyonic}. The last row is a control: off the locus the parametric form is not a solution, and the disagreement appears in the first decimal.}
\label{tab:limits}
\begin{tabularx}{\textwidth}{@{}LCCC@{}}
\toprule
{\bf limit or check} & {\bf quantity} & {\bf parametric form} & {\bf reference value} \\
\midrule
$a\to0$, dyonic Reissner--Nordstr\"om & $f(2.5)$ & $0.30599999999996$ & $0.30600000000000$ \\
$a\to0$, $\Lambda<0$, dyonic RN--AdS & $f(2.5)$ & $0.49349999999996$ & $0.49350000000000$ \\
$q\to0$, magnetic branch \eqref{eq:magnetic} & $f(2.5)$ & $0.25752568217642$ & $0.25752568217600$ \\
$p\to0$, electric branch, direct quadrature & $f(2.5)$ & $0.25753706432424$ & $0.25753706432424$ \\
$p,q\to0$, Schwarzschild & $f(2.5)$ & $0.20000000000010$ & $0.20000000000000$ \\
$p,q\to0$, $\Lambda<0$, Schwarzschild--AdS & $f(2.5)$ & $0.38750000000014$ & $0.38750000000000$ \\
dyonic, direct quadrature of \eqref{eq:master} & $f(1.898011)$ & $0.12943153318161$ & $0.12943153318161$ \\
dyonic, direct quadrature of \eqref{eq:master} & $f(0.601148)$ & $-0.93652843149713$ & $-0.93652843149713$ \\
large-$r$ series \eqref{eq:largeR} & $f(10)$ & $0.80692497189068$ & $0.80692497189068$ \\
near-core series \eqref{eq:core} & $f(10^{-4})$ & $9.610129\times10^{2}$ & $9.610121\times10^{2}$ \\
$J_{\infty}$, Beta form against $\Gamma$ form & $J_{\infty}(0.3)$ & $4.72389943102995$ & $4.72389943102996$ \\
$a\to0$ temperature, dyonic RN & $T(1.7)$ & $0.04016937297662$ & $0.04016937297658$ \\
$a\to0$ photon sphere, RN & $r_{\rm ph}$ & $2.73693518392214$ & $2.73693168768530$ \\
$a\to0$ shadow radius, RN & $b_{\rm ph}$ & $4.85869419620093$ & $4.85869419324685$ \\
control, off the locus, $b\neq a/2$ & $f(1.2)$ & $-0.21702364$ & $-0.24489820$ \\
\bottomrule
\end{tabularx}
\end{table}

Away from the locus the standard recourse is an expansion in the couplings, and it is worth recording what that expansion gives and where the exact solution supersedes it. Writing $F_{rt}=q/r^{2}+\delta$ in \eqref{eq:cubic} and retaining first order in $a$ and $b$,
\begin{equation}
F_{rt}\simeq\frac{q}{r^{2}}+\frac{4qp^{2}(a-2b)-4aq^{3}}{r^{6}} .
\label{eq:Fpert}
\end{equation}
In the integrand of \eqref{eq:master} the quartic term needs only the Maxwell field, since it carries an explicit factor of $a$, and the two contributions combine to
\begin{equation}
\int_{r}^{\infty}\dd r'\left[2a\,r'^{2}F_{rt}^{4}+q\,F_{rt}\right]
\simeq\frac{q^{2}}{r}+\frac{2q^{2}}{5r^{5}}\left[2p^{2}(a-2b)-aq^{2}\right],
\label{eq:Ipert}
\end{equation}
so that the lapse function reads
\begin{equation}
f\simeq1-\frac{2m}{r}-\frac{\Lambda r^{2}}{3}+\frac{p^{2}+q^{2}}{r^{2}}
+\frac{4p^{2}q^{2}(a-2b)-2aq^{4}-2ap^{4}}{5r^{6}} .
\label{eq:fpert}
\end{equation}
This holds for arbitrary couplings but only while $aq^{2}/r^{4}\ll1$ and $|a-2b|p^{2}/r^{4}\ll1$, conditions that fail near the horizon and fail badly near the origin, which is where the horizon phases of Sec.~\ref{isec3a} and the core structure of Sec.~\ref{isec3b} reside.

Three independent checks fix \eqref{eq:fpert}. At $q=0$ it returns the exact magnetic solution \eqref{eq:magnetic} rather than an approximation to it. At $a=0$ it reproduces the small-$b$ expansion of the exact quasitopological dyonic solution, whose bracket of logarithms and arctangents expands as $8\alpha/r-\tfrac{32}{5}\alpha^{5}/r^{5}$ with $\alpha^{4}=2bp^{2}$, leaving the correction $-8bp^{2}q^{2}/5r^{6}$; the residual against the exact solution scales as $b^{2}$. At $p=0$ it reproduces the large-$r$ expansion of the exact electric branch \citep{Luo2026}.

On the locus $b=a/2$ the cross term in \eqref{eq:fpert} vanishes identically and the perturbative lapse function collapses to
\begin{equation}
f\simeq1-\frac{2m}{r}-\frac{\Lambda r^{2}}{3}+\frac{p^{2}+q^{2}}{r^{2}}-\frac{2a\left(p^{4}+q^{4}\right)}{5r^{6}} ,
\label{eq:fpertlocus}
\end{equation}
which is precisely the first-order truncation of the exact solution \eqref{eq:dyonic}, as the large-$r$ expansion in Appendix~\ref{app:A} shows. The residual between the two scales as $a^{2}$. The exact solution therefore does not replace \eqref{eq:fpert} so much as complete it at one point of coupling space, supplying the all-orders resummation against which the perturbative treatment elsewhere can be calibrated.

Equation \eqref{eq:fpertlocus} is symmetric under $p\leftrightarrow q$. That symmetry is not an accident of the truncation and not exact either. The exact solution \eqref{eq:dyonic} is invariant under the interchange of the two charges through first order in $a$, and the leading asymmetry is second order, as the $r^{-10}$ term of \eqref{eq:largeR} makes plain: it carries $q^{6}$ with no magnetic counterpart. Evaluating the difference $f(r;p,q)-f(r;q,p)$ at $r=3$, $m=1$, $p=0.3$ and $q=0.6$ gives $1.3801\times10^{-6}a^{2}$ at $a=0.05$ and $1.3625\times10^{-6}a^{2}$ at $a=0.4$, a coefficient stable to better than two percent across a factor of eight in the coupling. The locus is therefore duality symmetric to the accuracy at which any perturbative treatment of this theory works, and the exact solution is what exposes where that symmetry fails.

\begin{figure}[!htbp]
\centering
\includegraphics[width=0.70\textwidth]{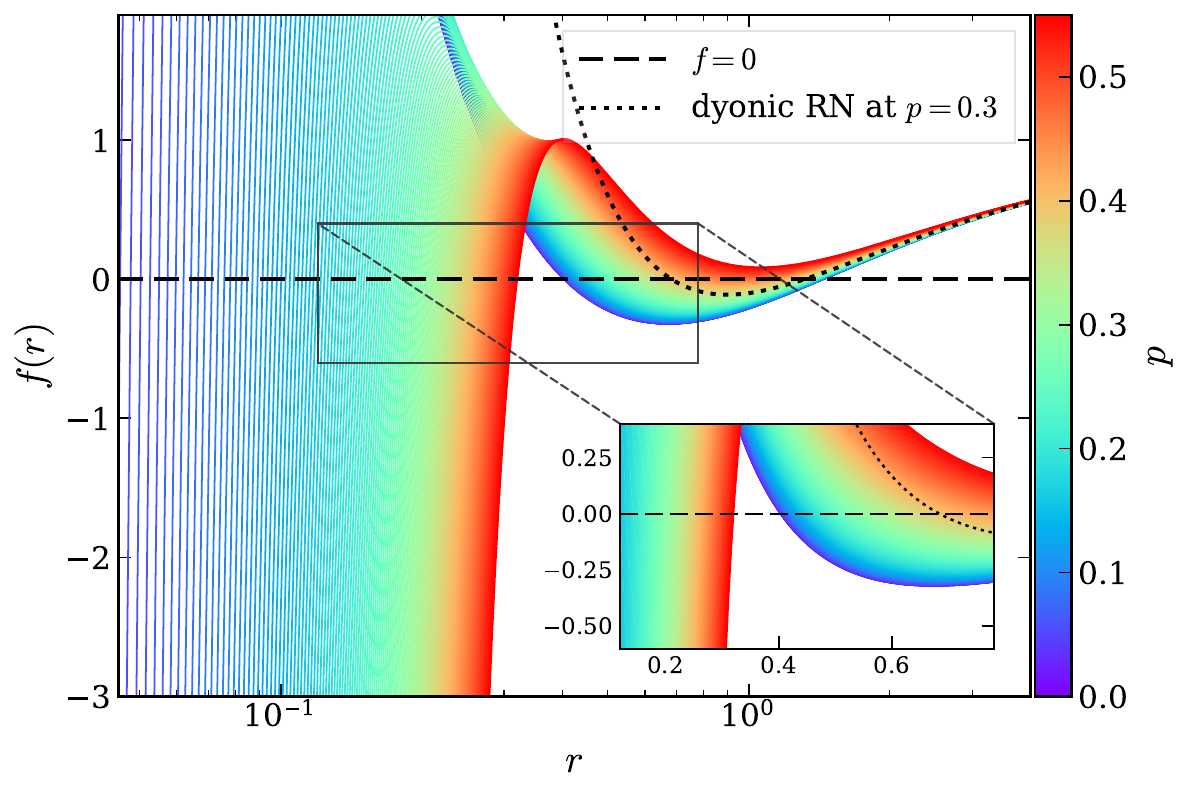}
\caption{Lapse function \eqref{eq:dyonic} at $a=0.1$, $q=0.9$, $m=1$ and $\Lambda=0$, drawn as a continuous family in the magnetic charge. The dotted curve is the dyonic Reissner--Nordstr\"om lapse function at $p=0.3$, which the exact solution approaches from below at large radius and departs from inside the photon sphere. The inset resolves the region where the inner roots appear and disappear.}
\label{fig:lapse}
\end{figure}

The behaviour of \eqref{eq:dyonic} across the whole magnetic range is shown in Fig.~\ref{fig:lapse}, where the family of lapse functions at fixed $a$, $q$ and $m$ is drawn as a continuous ribbon in $p$. Two things are visible at once. Far from the centre the curves collapse onto the dyonic Reissner--Nordstr\"om form, since the corrections enter at $r^{-6}$; near the centre they separate sharply, because the magnetic term drives $f$ to $-\infty$ for any $p\neq0$ while the electric sector alone leaves a finite self-energy and drives $f$ to $+\infty$. The inset resolves the crossing region, where the number of roots changes.

\subsection{Horizon structure and the three-horizon window}\label{isec3a}

Horizons are the roots of \eqref{eq:dyonic}, and because the magnetic sector contributes a term falling as $r^{-6}$ while the electric sector softens near the origin, the two compete rather than add. The result is a window in which three horizons exist, bounded above and below in each charge separately. Multihorizon geometries are known to occur in several nonlinear theories \citep{Gao2021,Nashed2021}, and what distinguishes the present case is which charge opens the window and which closes it.

Table~\ref{tab:horizons} shows the effect at fixed $a$ and $m$ with $\Lambda=0$. With the magnetic charge switched off the geometry has the two horizons of the electric branch. A small magnetic charge opens a third, because the $-2ap^{4}/5r^{6}$ term overwhelms the softened electric core and returns $f$ to $-\infty$; a larger one closes the window and leaves a single horizon. Holding the magnetic charge fixed and varying the electric charge does the same from the other side, since the inner pair of the electric branch exists only while the effective core mass $M_{\rm eff}=m-\tfrac12 q^{3/2}J_{\infty}$ is negative, which requires $q$ above a threshold, and since a large enough $q$ eventually swallows the outer root as well.

\begin{table}[!htbp]
\centering
\caption{Horizon radii of the dyonic solution \eqref{eq:dyonic} at $a=0.1$, $m=1$ and $\Lambda=0$, showing the three-horizon window and its closure in either charge. The two-horizon entry sits exactly on the axis $p=0$; any nonzero magnetic charge inside the window produces three roots.}
\label{tab:horizons}
\begin{tabularx}{\textwidth}{@{}>{\hsize=0.55\hsize}C>{\hsize=0.55\hsize}C>{\hsize=1.95\hsize}L>{\hsize=0.95\hsize}L@{}}
\toprule
{\bf $p$} & {\bf $q$} & {\bf horizon radii} & {\bf phase} \\
\midrule
$0.0$ & $0.9$ & $0.403117,\ 1.442296$ & two horizons, electric branch \\
$0.1$ & $0.9$ & $0.079690,\ 0.420135,\ 1.431037$ & three horizons \\
$0.3$ & $0.9$ & $0.205052,\ 0.554139,\ 1.328216$ & three horizons \\
$0.5$ & $0.9$ & $0.300733$ & single horizon \\
$0.3$ & $0.6$ & $1.742017$ & single horizon \\
$0.3$ & $1.1$ & $0.173539$ & single horizon \\
\bottomrule
\end{tabularx}
\end{table}

\begin{figure}[!htbp]
\centering
\includegraphics[width=0.70\textwidth]{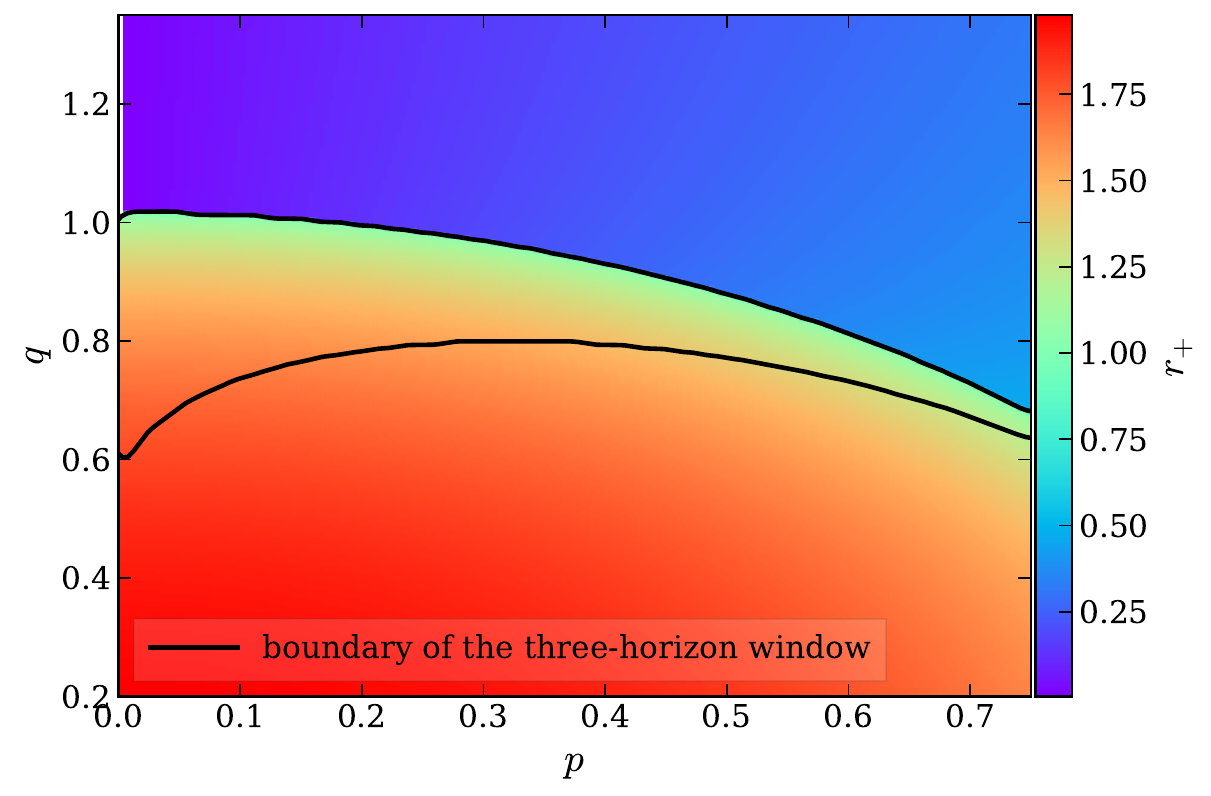}
\caption{Outer horizon radius of \eqref{eq:dyonic} over the charge plane at $a=0.1$, $m=1$ and $\Lambda=0$. The solid curve bounds the region in which three horizons exist. The axis $p=0$ inside that band is the two-horizon line of the purely electric branch, and any nonzero magnetic charge there opens a third root. The colour field is continuous across the boundary, so the outer horizon carries no imprint of the interior phase.}
\label{fig:phase}
\end{figure}

The full extent of the window is mapped in Fig.~\ref{fig:phase}, where the outer horizon radius is drawn as a continuous field over the charge plane and the phase boundaries are overlaid on it. The three-horizon region is a band, open at $p=0$ and closing at large $p$, and bounded in $q$ from both sides. Its lower boundary in $q$ is where the electric core mass changes sign; its upper boundary is where the outer and middle roots merge and annihilate. This is not the phase reported for the magnetic and dyonic Euler--Heisenberg solutions, where the three-horizon configuration is driven by the magnetic sector alone and is absent from the electric branch and from dyonic Reissner--Nordstr\"om \citep{Luo2026}. Here the innermost root is magnetic in origin while the middle root is electric, so the configuration requires both charges at once and is destroyed by an excess of either. The outer horizon radius itself varies smoothly across the whole plane and carries no visible signature of the transition, which is worth keeping in mind for Sec.~\ref{isec6}: an external observer measuring only the outer horizon or the shadow would not know which phase the interior is in.

\subsection{Curvature, the character of the core, and the energy conditions}\label{isec3b}

The two sectors that compete for the horizon structure also compete for the singularity, and they do so with different powers. This section separates them, identifies a tuning at which the electric singularity is softened well below the Schwarzschild strength, and then derives the energy density and the pressures in closed form, from which the admissible region of parameter space follows.

For the metric \eqref{eq:metric} the Ricci and Kretschmann scalars are
\begin{equation}
R=-f''-\frac{4f'}{r}+\frac{2(1-f)}{r^{2}},
\qquad
K=\left(f''\right)^{2}+\frac{4\left(f'\right)^{2}}{r^{2}}+\frac{4(1-f)^{2}}{r^{4}} ,
\label{eq:invariants}
\end{equation}
with $f'$ and $f''$ supplied in closed form by \eqref{eq:fprime} and \eqref{eq:fpprime}. Near the centre three regimes occur. For $p\neq0$ the magnetic term dominates and
\begin{equation}
K\simeq\frac{7648\,a^{2}p^{8}}{25\,r^{16}} ,
\label{eq:Kmag}
\end{equation}
a divergence far stronger than the $r^{-8}$ of Reissner--Nordstr\"om, which we have confirmed to six figures at $a=0.3$, $p=q=0.6$, where $Kr^{16}$ settles at $0.4624453140$ against the predicted $0.4624453140$. For $p=0$ the electric sector contributes no Coulomb divergence, the expansion \eqref{eq:core} leaves a term linear in $1/r$ with the self-energy subtracted, and
\begin{equation}
K\simeq\frac{48M_{\rm eff}^{2}}{r^{6}},\qquad M_{\rm eff}=m-\tfrac12 q^{3/2}J_{\infty} ,
\label{eq:Kel}
\end{equation}
which is Schwarzschild strength with a shifted mass. The third regime is the interesting one. If the mass is tuned to the electric self-energy, $M_{\rm eff}=0$, the leading term of \eqref{eq:core} disappears and the subleading $r^{-2/3}$ piece takes over, giving
\begin{equation}
K\simeq\frac{568\,q^{8/3}}{2^{10/3}a^{2/3}}\,\frac{1}{r^{16/3}} ,
\label{eq:Krel}
\end{equation}
a singularity weaker than Schwarzschild's. At $a=0.3$ and $q=0.6$ the tuning is $m=1.0977350295$ and the combination $Kr^{16/3}$ approaches $32.2035820$ against the predicted $32.2035820$, stable over six decades in radius. Relegated singularities of this kind are familiar from Born--Infeld type theories, where the finite self-energy of the source weakens the divergence without removing it \citep{Yang2022,Yang2023}. What is new is that here the tuning is a single algebraic condition on the mass, expressible in Beta functions through \eqref{eq:Jinf}, rather than a property of the Lagrangian.

\begin{figure}[!htbp]
\centering
\includegraphics[width=0.70\textwidth]{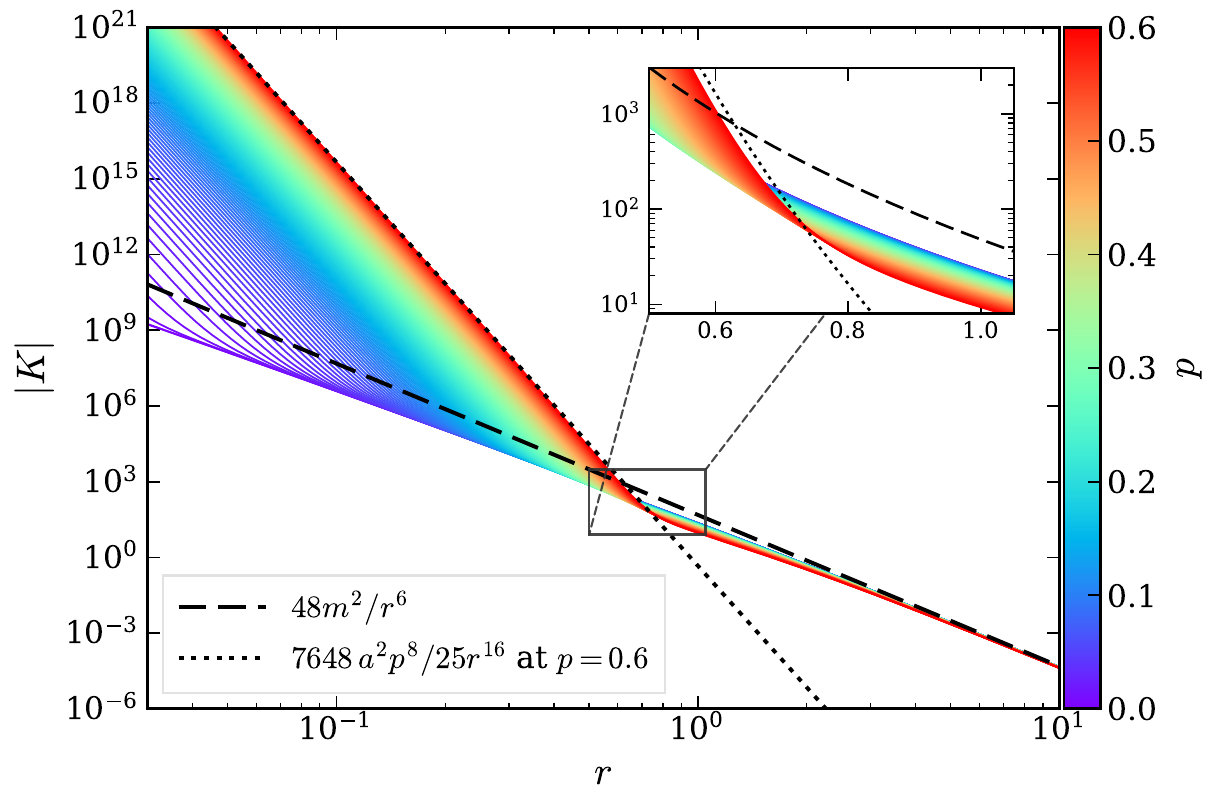}
\caption{Kretschmann scalar of \eqref{eq:dyonic} at $a=0.3$, $q=0.6$, $m=1$ and $\Lambda=0$, drawn as a continuous family in the magnetic charge. The dashed reference is the Schwarzschild law and the dotted one the magnetic core law \eqref{eq:Kmag} at $p=0.6$. Turning on the magnetic charge converts a Schwarzschild-strength singularity into one three powers stronger.}
\label{fig:kretschmann}
\end{figure}

Figure~\ref{fig:kretschmann} shows the transition between the first two regimes as the magnetic charge is turned on. The purely electric curve tracks the Schwarzschild reference over several decades; any nonzero $p$ eventually peels away and steepens to the $r^{-16}$ law, and the radius at which it does so moves outward as $p$ grows.

\begin{figure}[!htbp]
\centering
\includegraphics[width=0.70\textwidth]{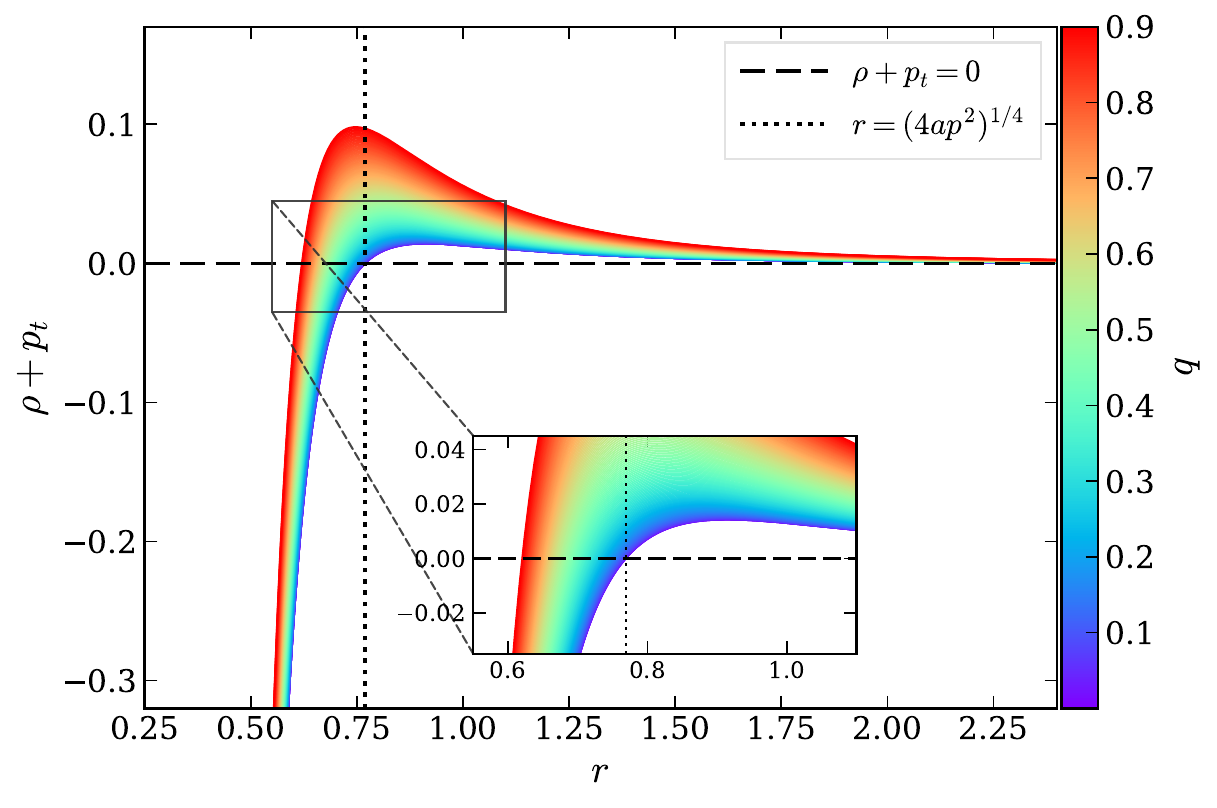}
\caption{The combination $\rho+p_{t}$ at $a=0.35$ and $p=0.5$, drawn as a continuous family in the electric charge. The vertical reference marks the purely magnetic boundary $r=(4ap^{2})^{1/4}$. The zero crossing moves inward as the electric charge grows, so the dominant energy condition holds over a larger region than the magnetic bound alone predicts. The inset resolves the crossing.}
\label{fig:dec}
\end{figure}

The source that supports \eqref{eq:dyonic} is anisotropic with $p_{r}=-\rho$, as it must be for a metric with $g_{tt}g_{rr}=-1$. Carrying the change of variable \eqref{eq:rofw} through the Einstein equations gives the energy density and the tangential pressure in closed form,
\begin{equation}
8\pi\rho=\frac{p^{2}}{r^{4}}-\frac{2ap^{4}}{r^{8}}+w^{4}\left(1+6aw^{4}\right),
\qquad
8\pi p_{t}=\frac{p^{2}}{r^{4}}-\frac{6ap^{4}}{r^{8}}+w^{4}\left(1+2aw^{4}\right),
\label{eq:rhopt}
\end{equation}
both of which reduce to $(p^{2}+q^{2})/8\pi r^{4}$ in the Maxwell limit, as the dyonic Reissner--Nordstr\"om source requires. Two combinations decide the energy conditions. The difference is
\begin{equation}
8\pi\left(\rho-p_{t}\right)=\frac{4ap^{4}}{r^{8}}+4aw^{8} ,
\label{eq:rhominus}
\end{equation}
which is nonnegative everywhere for $a>0$, so the dominant energy condition reduces to the single requirement $\rho+p_{t}\geq0$. Using $w^{2}(1+4aw^{4})=q/r^{2}$ to eliminate the bracket, that requirement becomes
\begin{equation}
\frac{p^{2}}{r^{4}}-\frac{4ap^{4}}{r^{8}}+\frac{q\,w^{2}}{r^{2}}\ \geq\ 0
\qquad\Longleftrightarrow\qquad
r^{4}-4ap^{2}+\frac{q\,w^{2}r^{6}}{p^{2}}\ \geq\ 0 .
\label{eq:DEC}
\end{equation}
At $q=0$ this is exactly the magnetic bound $r^{4}>4ap^{2}$ of Refs.~\citep{Croney2025,Croney2026}. The new term is manifestly positive, so the electric charge relaxes the bound: the radius below which the dominant energy condition fails moves inward as $q$ grows. Numerically, at $a=0.35$ and $p=0.5$ the boundary sits at $r=0.769138$ for $q=0$, exactly at $(4ap^{2})^{1/4}=0.769161$, and moves to $0.744271$, $0.701671$, $0.663979$ and $0.633217$ for $q=0.2$, $0.4$, $0.6$ and $0.8$. Over the same range the outer horizon shrinks from $1.866442$ to $1.355273$, so the violating region stays well inside the horizon throughout, and every configuration in Table~\ref{tab:horizons} is admissible in the exterior. Figure~\ref{fig:dec} displays the effect as a continuous family.

\section{The electric limit, the Smarr integral and the extended thermodynamics}\label{isec4}

Setting $p=0$ in \eqref{eq:dyonic} places us on the first locus of \eqref{eq:loci}, where the solution is already known. We record the agreement, which serves as an independent check of the construction, and then show that the parametrization settles a question this sector has left open.

At $p=0$ and $\Lambda=0$ the metric function reduces to
\begin{equation}
f=1-\frac{2m-q^{3/2}J(w)}{r(w)} ,
\label{eq:electric}
\end{equation}
which is the purely electric black hole obtained in Ref.~\citep{Luo2026} by a different route, in which the Cardano root is parametrized by an auxiliary variable rather than by the field strength itself. The two agree in every quantity we have compared. The constant $J_{\infty}$ of Appendix~\ref{app:A} equals their $\Gamma$-function expression to fourteen figures at $a=0.1$, $0.3$ and $1.0$, the coefficient of the subleading $r^{-2/3}$ term at the origin is identical, and the large-$r$ expansions coincide term by term. Since $b$ is absent from \eqref{eq:electric}, the same solution serves the Euler--Heisenberg and truncated Born--Infeld theories alike, a point already used in the geodesic and lensing studies of the electric branch \citep{Amaro2020,Amaro2022,Breton2022}.

The Komar mass relation for the purely electric configuration reads
\begin{equation}
M=\frac{1}{2}T_{\rm H}A_{\rm H}+\int_{\rH}^{\infty}\left(r^{2}V'^{2}+2a\,r^{2}V'^{4}\right)\dd r ,
\label{eq:smarrint}
\end{equation}
and the remaining integral has been reported as having no closed form in general, since $V'$ is itself fixed by the cubic \citep{Luo2026,Gulin2018}. In the variable of \eqref{eq:rofw} it does have one. With $V'=-w^{2}$ the integrand becomes algebraic and
\begin{equation}
\int_{\rH}^{\infty}\left(r^{2}V'^{2}+2a\,r^{2}V'^{4}\right)\dd r
=q^{3/2}\int_{0}^{\wH}\frac{(1+2as^{4})(1+12as^{4})}{(1+4as^{4})^{5/2}}\,\dd s ,
\label{eq:smarrclosed}
\end{equation}
which evaluates to the three hypergeometric terms of \eqref{eq:smarrhyp}. The Smarr relation therefore closes analytically for the electric branch. At $a=0.3$, $q=0.6$ and $m=1$ the horizon lies at $\rH=1.800923$ with $T_{\rm H}=0.0393764$, the integral evaluates to $0.42518137464781$ by direct quadrature and to $0.42518137464781$ through \eqref{eq:smarrhyp}, and half the product $T_{\rm H}A_{\rm H}$ plus that value returns the mass to twelve decimal places. General statements about when a Smarr formula exists in nonlinear electrodynamics, and how it must be generalised, have been given in Refs.~\citep{Gulin2018,Zhang2018,Balart2017,Breton2005,Pereira2014}; what \eqref{eq:smarrclosed} adds is that in this particular theory the obstruction was one of variables rather than of principle.

The large-$r$ form of the dyonic lapse function is what enters weak-field constraints, lensing at large impact parameter and the boundary conditions of numerical integrations, so its coefficients matter. Evaluating \eqref{eq:fpert} at the Euler--Heisenberg couplings $a=\epsilon$, $b=7\epsilon/4$ and writing $\beta=p/q$ gives
\begin{equation}
f\simeq1-\frac{2m}{r}+\frac{\left(1+\beta^{2}\right)q^{2}}{r^{2}}
-\frac{2\epsilon\left(1+5\beta^{2}+\beta^{4}\right)q^{4}}{5r^{6}} .
\label{eq:asymEH}
\end{equation}
This differs from the expression quoted in Ref.~\citep{Luo2026} for the same quantity, which carries a term odd in the magnetic charge and omits the factor of five in the denominator. The form \eqref{eq:asymEH} is even in $p$, as any expansion in the invariants must be, it reduces at $\beta=0$ to the $-2\epsilon q^{4}/5r^{6}$ of their own exact electric solution, and it agrees with the exact solution of Sec.~\ref{isec3} at the locus and with the exact quasitopological solution at $a=0$. We therefore take the discrepancy to be typographical, and record \eqref{eq:asymEH} because the boundary conditions of a dyonic integration in this sector are set by it.
Turning now to the thermodynamics, the solution has five independent parameters once the cosmological constant is promoted to a pressure, and all five appear in the first law. This section evaluates the conjugates in closed form, verifies the first law and the Smarr formula numerically, and then examines the local stability and the extended phase space, where the coupling turns out to weaken and eventually destroy the van der Waals behaviour that the charged anti-de Sitter black hole is known for. Throughout, the horizon is the outermost root of \eqref{eq:dyonic}, at $w=\wH$, and the mass parameter follows from $f(\rH)=0$. Writing the pressure as $P=-\Lambda/8\pi$ and the thermodynamic volume as $V=\tfrac{4}{3}\pi \rH^{3}$,
\begin{equation}
\Ms=\frac{\rH}{2}+PV+\frac{p^{2}}{2\rH}-\frac{ap^{4}}{5\rH^{5}}+\frac{q^{3/2}}{2}J(\wH),
\qquad
S=\pi \rH^{2} .
\label{eq:mass}
\end{equation}
The entropy obeys the area law because the nonlinearity sits in the matter sector and not in the gravitational action. Evaluating the surface gravity through the chain rule and eliminating the mass gives a temperature that is elementary despite the hypergeometric mass,
\begin{equation}
4\pi T=\frac{1}{\rH}+8\pi P\rH-\frac{p^{2}}{\rH^{3}}+\frac{2ap^{4}}{\rH^{7}}
-\frac{q\,\wH^{2}\left(1+6a\wH^{4}\right)}{\rH\left(1+4a\wH^{4}\right)} ,
\label{eq:temperature}
\end{equation}
which returns the dyonic Reissner--Nordstr\"om--anti-de Sitter temperature at $a=0$, where $\wH^{2}=q/\rH^{2}$ and the last term collapses to $q^{2}/\rH^{3}$.

\subsection{The first law, the conjugate of the coupling, and the Smarr formula}\label{isec4a}

Both charges, the nonlinear coupling and the pressure are independent parameters of \eqref{eq:dyonic}, so the first law carries five terms,
\begin{equation}
\dd \Ms=T\,\dd S+\Psi\,\dd p+\Phi\,\dd q+A\,\dd a+V\,\dd P ,
\label{eq:firstlaw}
\end{equation}
with $\Psi$, $\Phi$ and $A$ the partial derivatives of \eqref{eq:mass} at fixed remaining variables. Treating a coupling constant of the matter Lagrangian as a thermodynamic variable is standard once the Smarr relation is required to hold, and has been used for Born--Infeld and for quasitopological electrodynamics \citep{Gunasekaran2012,Kuang2018,Barrientos2022}. The magnetic potential is elementary,
\begin{equation}
\Psi=\frac{p}{\rH}-\frac{4ap^{3}}{5\rH^{5}} ,
\label{eq:Psi}
\end{equation}
and the electric potential follows from differentiating $J(\wH)$ at fixed horizon radius, which moves $\wH$ through the constitutive relation,
\begin{equation}
\Phi=\frac{3}{4}\sqrt{q}\,J(\wH)+\frac{\sqrt{q}\,\wH\left(1+6a\wH^{4}\right)}{4\left(1+4a\wH^{4}\right)^{3/2}} .
\label{eq:Phi}
\end{equation}
At $a=0$ both reduce to the Coulomb values $p/\rH$ and $q/\rH$. The conjugate of the coupling requires the derivative of $J$ with respect to $a$ at fixed $w$, which is itself a hypergeometric quadrature; writing it as $K(w,a)$ and combining it with the induced motion of $\wH$,
\begin{equation}
A=-\frac{p^{4}}{5\rH^{5}}
+\frac{q^{3/2}}{2}\left[K(\wH,a)-\frac{2\wH^{5}\left(1+6a\wH^{4}\right)}{\left(1+4a\wH^{4}\right)^{5/2}}\right],
\qquad
K(w,a)=\int_{0}^{w}\frac{s^{4}\left(8+36as^{4}-144a^{2}s^{8}\right)}{\left(1+4as^{4}\right)^{7/2}}\,\dd s ,
\label{eq:Acoup}
\end{equation}
whose closed form is \eqref{eq:Khyp}. The field equations are invariant under $r\to\lambda r$, $(m,p,q)\to\lambda(m,p,q)$, $a\to\lambda^{2}a$, $\Lambda\to\lambda^{-2}\Lambda$, so $\Ms$ is homogeneous of degree one with $S$ of degree two, $p$ and $q$ of degree one, $a$ of degree two and $P$ of degree minus two, and Euler's theorem gives
\begin{equation}
\Ms=2TS+\Psi p+\Phi q+2Aa-2PV .
\label{eq:smarr}
\end{equation}
Table~\ref{tab:thermo} verifies both relations. The Smarr formula holds to the last digit carried, and the first law holds along a generic direction in the five-dimensional parameter space to the accuracy of the finite difference used to test it. The $a\to0$ limit reproduces the dyonic Reissner--Nordstr\"om--anti-de Sitter results, the $q\to0$ limit the magnetic ones, and the $\Lambda\to0$ limit removes the last term of \eqref{eq:smarr}; all four limits were checked separately and agree to the same accuracy.

\begin{table}[!htbp]
\centering
\caption{Closed-form potentials, the Smarr formula \eqref{eq:smarr} and the first law \eqref{eq:firstlaw}, verified at three points of parameter space. The last column is the relative residual of the first law along the generic direction $(\dd \rH,\dd a,\dd p,\dd q,\dd P)\propto(3.1,2.3,1.7,-2.9,0.011)\times10^{-6}$, limited by the finite difference and not by the closed forms.}
\label{tab:thermo}
\begin{tabularx}{\textwidth}{@{}CCCCCCCC@{}}
\toprule
{\bf $\rH$} & {\bf $a$} & {\bf $p$} & {\bf $q$} & {\bf $P$} & {\bf $\Ms$} & {\bf Smarr, Eq.~\eqref{eq:smarr}} & {\bf first law} \\
\midrule
$1.70$ & $0.30$ & $0.45$ & $0.55$ & $0.0045$ & $1.0905944792$ & $1.0905944792$ & $1.6\times10^{-7}$ \\
$2.40$ & $0.60$ & $0.20$ & $0.80$ & $0.0020$ & $1.4568876870$ & $1.4568876870$ & $1.1\times10^{-8}$ \\
$1.20$ & $0.15$ & $0.55$ & $0.30$ & $0$ & $0.7623434643$ & $0.7623434643$ & $4.4\times10^{-8}$ \\
\bottomrule
\end{tabularx}
\end{table}

\begin{figure}[!htbp]
\centering
\includegraphics[width=0.70\textwidth]{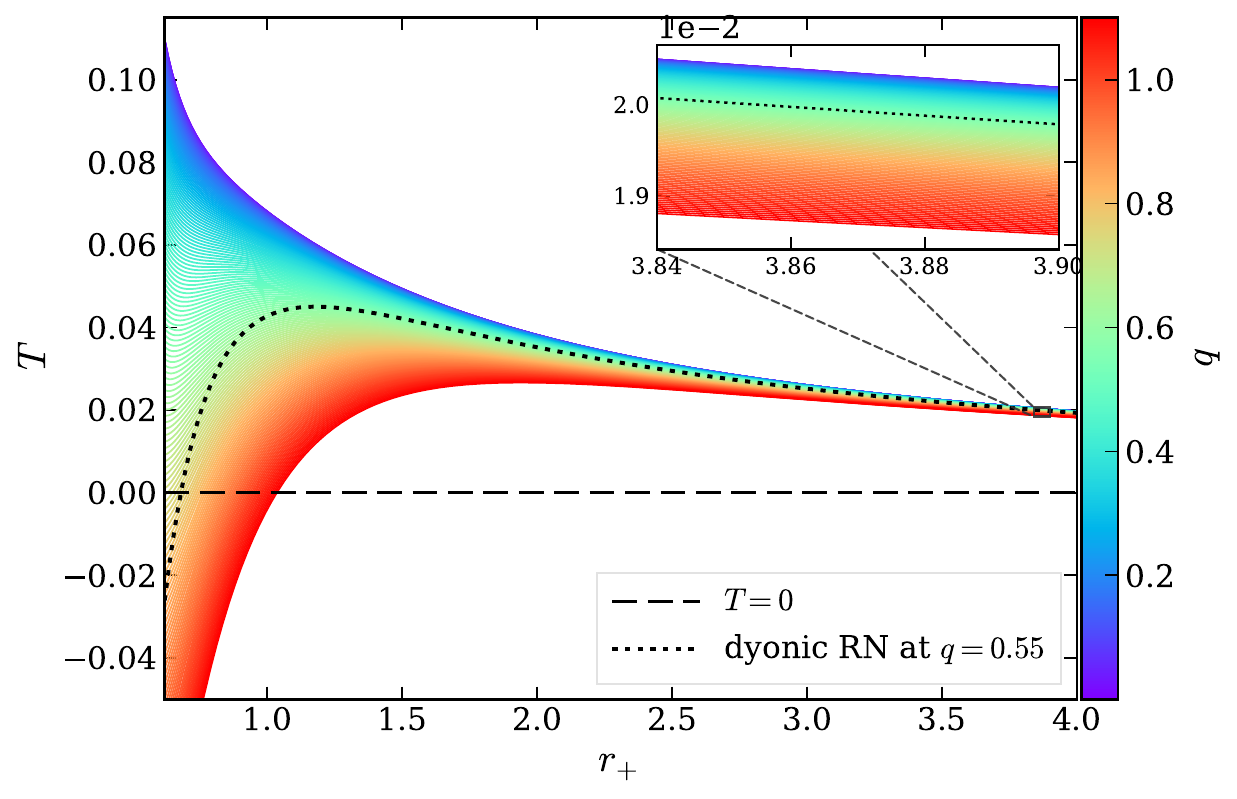}
\caption{Hawking temperature \eqref{eq:temperature} at $a=0.3$, $p=0.4$ and $\Lambda=0$, drawn as a continuous family in the electric charge. The dotted reference is the dyonic Reissner--Nordstr\"om temperature at $q=0.55$. The inset resolves the extremal root and the maximum, which is where the heat capacity of Fig.~\ref{fig:heatcapacity} changes sign.}
\label{fig:temperature}
\end{figure}

\begin{figure}[!htbp]
\centering
\includegraphics[width=0.70\textwidth]{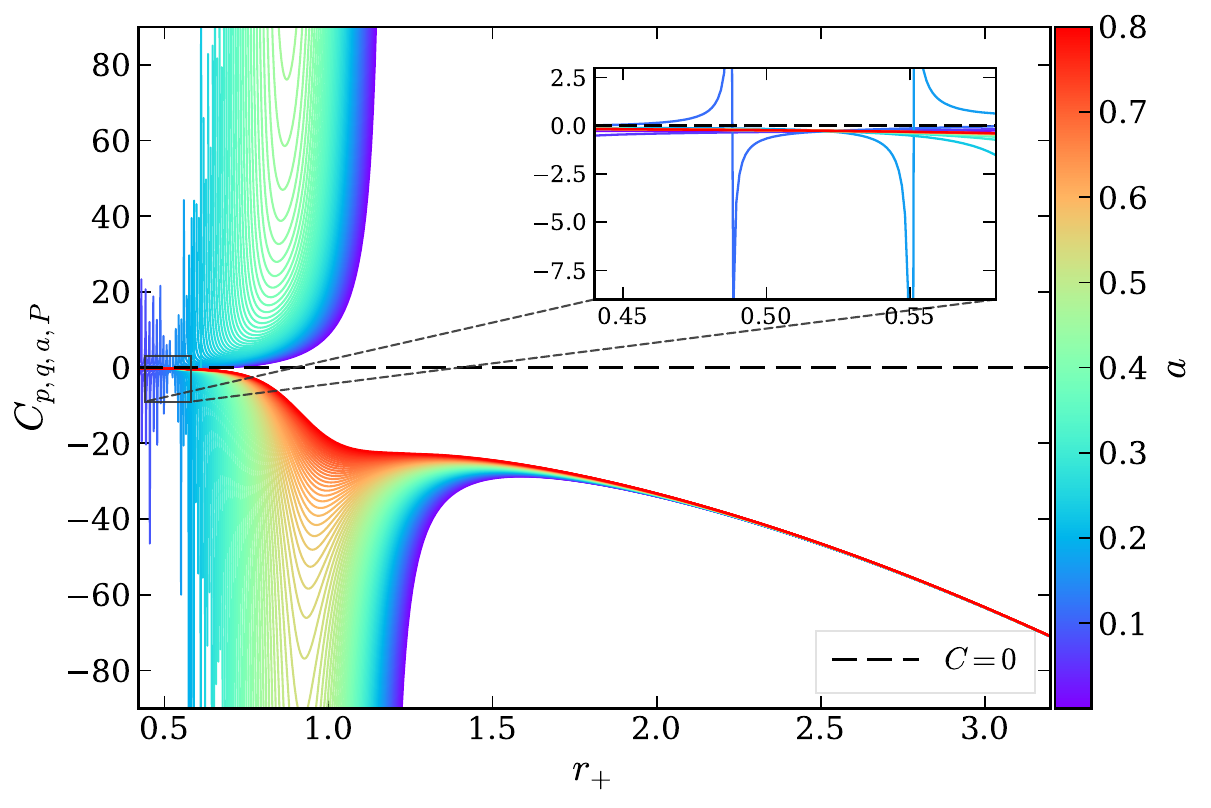}
\caption{Heat capacity \eqref{eq:heatcapacity} at $p=0.4$, $q=0.55$ and $\Lambda=0$, drawn as a continuous family in the nonlinear coupling. The divergence separates the small unstable branch from the large stable one and moves outward as the coupling grows.}
\label{fig:heatcapacity}
\end{figure}

The temperature itself is shown in Fig.~\ref{fig:temperature} as a continuous family in the electric charge. The pattern is the familiar one for charged black holes, with an extremal root at small radius, a maximum, and the Schwarzschild falloff beyond it, but the extremal radius drifts inward relative to the dyonic Reissner--Nordstr\"om reference because the softened electric core reduces the electrostatic contribution to the surface gravity. Local stability follows from the heat capacity at fixed charges, coupling and pressure,
\begin{equation}
C_{p,q,a,P}=T\left(\frac{\partial S}{\partial T}\right)_{p,q,a,P}
=\frac{2\pi \rH T}{\left(\partial T/\partial \rH\right)_{p,q,a,P}} ,
\label{eq:heatcapacity}
\end{equation}
plotted in Fig.~\ref{fig:heatcapacity}. It is negative on the small-radius branch, diverges where the temperature peaks, and is positive beyond, so the divergence separates a locally unstable branch from a stable one in the usual way. Increasing the coupling moves the divergence outward and widens the unstable branch, which is the thermodynamic counterpart of the statement in Sec.~\ref{isec3b} that the coupling softens the electric core.

\subsection{Equation of state and criticality}\label{isec4b}

With $\Lambda<0$ the temperature \eqref{eq:temperature} inverts to an equation of state,
\begin{equation}
P=\frac{T}{2\rH}-\frac{1}{8\pi \rH^{2}}+\frac{p^{2}}{8\pi \rH^{4}}-\frac{ap^{4}}{4\pi \rH^{8}}
+\frac{q\,\wH^{2}\left(1+6a\wH^{4}\right)}{8\pi \rH^{2}\left(1+4a\wH^{4}\right)} ,
\label{eq:eos}
\end{equation}
whose Maxwell limit is the charged anti-de Sitter equation of state with $p^{2}+q^{2}$ in place of the single charge. Writing $P=T/2\rH+g(\rH)$, the critical point is the root of $2g'+\rH g''=0$, after which $T_{c}=2\rH^{2}g'$ and $P_{c}$ follow directly. Table~\ref{tab:crit} collects the result. At $a\to0$ the critical ratio is $P_{c}v_{c}/T_{c}=3/8$ with $v=2\rH$, as it must be. Increasing the coupling lowers it monotonically, to $0.3183$ at $a=0.8$ for the purely electric configuration, so the fluid analogy survives but the universality of the van der Waals value does not. Departures from $3/8$ have been reported for Born--Infeld and other nonlinear sources \citep{Gunasekaran2012,Kruglov2022,Kruglov2022a,Wu2018}, and the present numbers fall in the same range.

\begin{figure}[!htbp]
\centering
\includegraphics[width=0.70\textwidth]{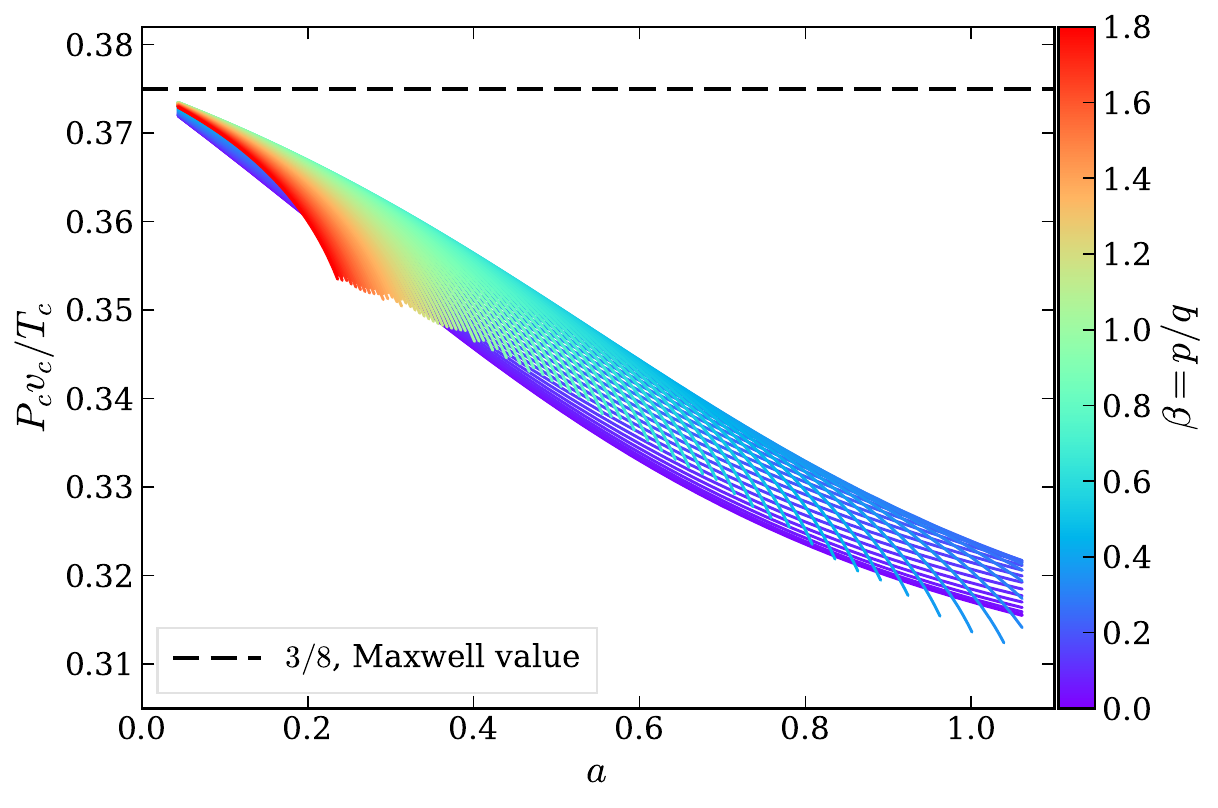}
\caption{Critical ratio of \eqref{eq:eos} against the nonlinear coupling, drawn as a continuous family in the charge ratio at fixed total charge $\sqrt{p^{2}+q^{2}}=0.55$. Every curve leaves the Maxwell value $3/8$ as soon as the coupling is switched on, and the departure is faster the larger the magnetic fraction.}
\label{fig:critratio}
\end{figure}

\begin{figure}[!htbp]
\centering
\includegraphics[width=0.70\textwidth]{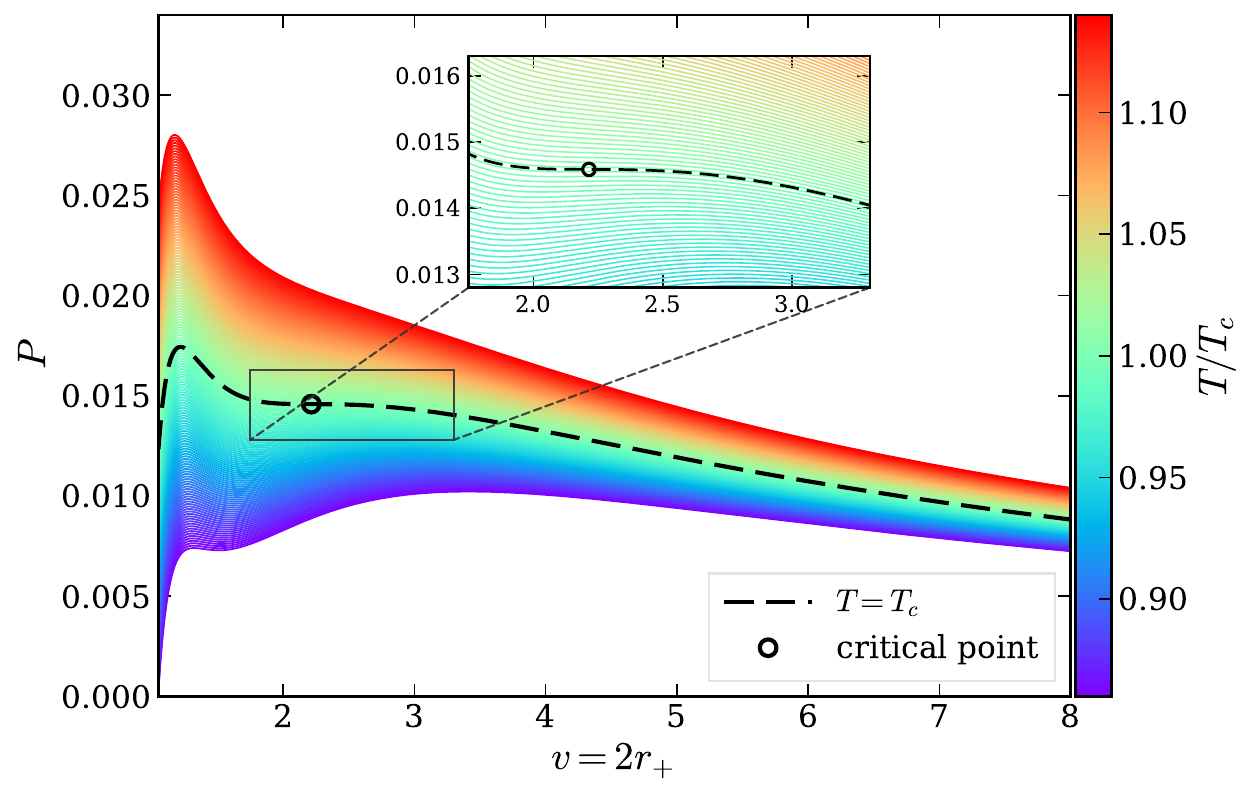}
\caption{Isotherms of \eqref{eq:eos} at $a=0.3$, $p=0.3$ and $q=0.4$, drawn as a continuous family in $T/T_{c}$. The dashed curve is the critical isotherm and the open circle the critical point. The inset resolves the oscillation that produces the first-order transition below $T_{c}$.}
\label{fig:isotherms}
\end{figure}

\begin{figure}[!htbp]
\centering
\includegraphics[width=0.70\textwidth]{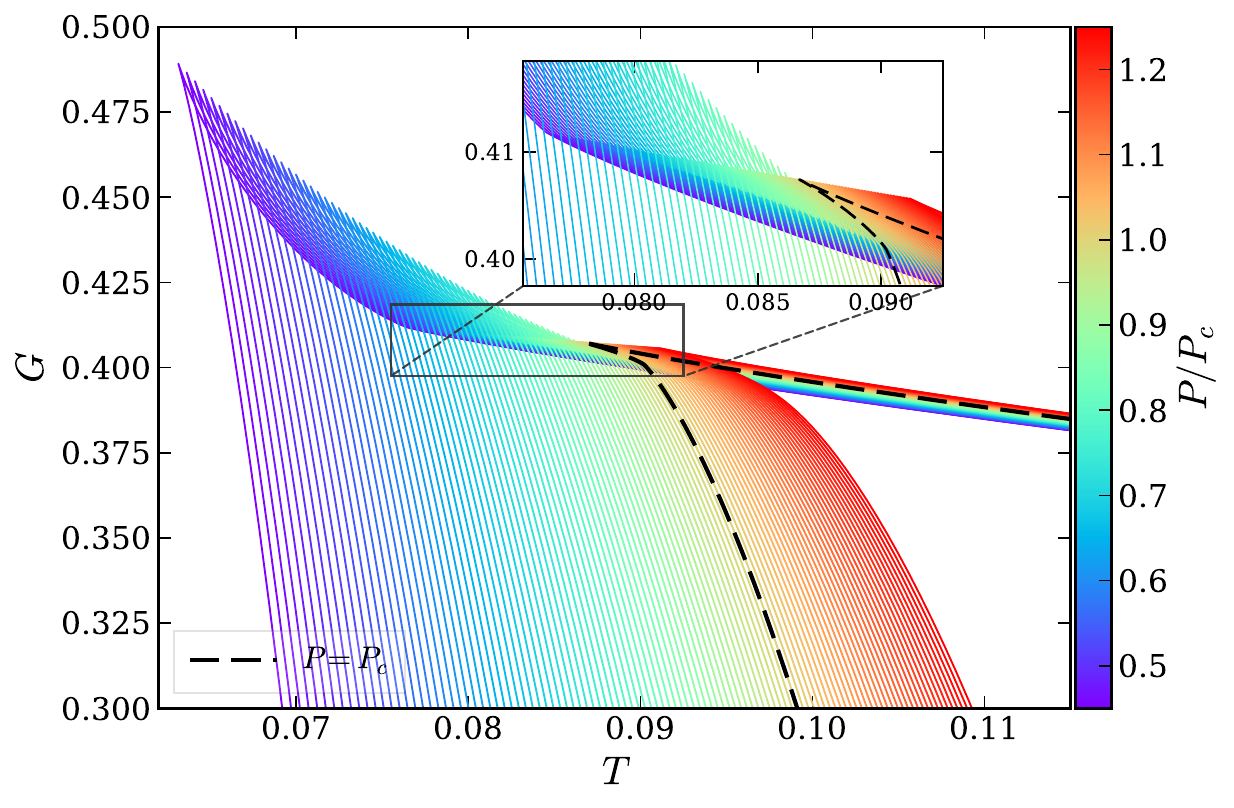}
\caption{Gibbs free energy $G=\Ms-TS$ at $a=0.3$, $p=0.3$ and $q=0.4$, drawn as a continuous family in $P/P_{c}$. The dashed curve is the critical pressure, at which the swallowtail closes. The inset resolves the crossing where the small and large branches exchange dominance.}
\label{fig:gibbs}
\end{figure}

More striking is that the critical point can disappear altogether. For the purely magnetic configuration at $p=0.5$ it exists at $a=0.14$ and is gone by $a=0.16$; for $p=0.3$, $q=0.4$ it survives to $a=0.45$ and is gone by $a=0.50$; for the purely electric configuration at $q=0.5$ it persists past $a=0.8$. The magnetic sector is therefore what destroys the transition, which is consistent with \eqref{eq:eos}, where the magnetic coupling enters through $-ap^{4}/4\pi \rH^{8}$ with a steep negative power while the electric contribution is bounded. Figure~\ref{fig:critratio} shows the ratio as a continuous family in the charge ratio $\beta=p/q$ at fixed total charge, and Figs.~\ref{fig:isotherms} and \ref{fig:gibbs} show the isotherms and the Gibbs free energy at a representative point inside the region where criticality survives. The swallowtail closes at $P=P_{c}$ in the standard way, so the transition is first order below the critical pressure and the small and large branches exchange dominance at the crossing.

\begin{table}[!htbp]
\centering
\caption{Critical point of \eqref{eq:eos}. The Maxwell row reproduces $P_{c}v_{c}/T_{c}=3/8$ exactly. Increasing the coupling lowers the ratio and, in the magnetically dominated configurations, eventually removes the critical point.}
\label{tab:crit}
\begin{tabularx}{\textwidth}{@{}CCCCCCC@{}}
\toprule
{\bf $a$} & {\bf $p$} & {\bf $q$} & {\bf $\rH^{c}$} & {\bf $T_{c}$} & {\bf $P_{c}$} & {\bf $P_{c}v_{c}/T_{c}$} \\
\midrule
$\to0$ & $0$ & $0.5$ & $1.224745$ & $0.086633$ & $0.013263$ & $0.375000$ \\
$0.05$ & $0$ & $0.5$ & $1.192594$ & $0.087619$ & $0.013619$ & $0.370731$ \\
$0.15$ & $0$ & $0.5$ & $1.126980$ & $0.089733$ & $0.014406$ & $0.361848$ \\
$0.30$ & $0$ & $0.5$ & $1.028644$ & $0.093253$ & $0.015789$ & $0.348327$ \\
$0.50$ & $0$ & $0.5$ & $0.909788$ & $0.098476$ & $0.018007$ & $0.332720$ \\
$0.80$ & $0$ & $0.5$ & $0.773521$ & $0.106879$ & $0.021960$ & $0.317864$ \\
$0.05$ & $0.3$ & $0.4$ & $1.207401$ & $0.087161$ & $0.013453$ & $0.372705$ \\
$0.30$ & $0.3$ & $0.4$ & $1.107397$ & $0.090189$ & $0.014585$ & $0.358173$ \\
$0.45$ & $0.3$ & $0.4$ & $1.004128$ & $0.092708$ & $0.015603$ & $0.338078$ \\
$0.05$ & $0.5$ & $0$ & $1.188387$ & $0.087699$ & $0.013649$ & $0.369918$ \\
$0.14$ & $0.5$ & $0$ & $1.039155$ & $0.090756$ & $0.014840$ & $0.339896$ \\
\bottomrule
\end{tabularx}
\end{table}

\section{Photon cones and vacuum birefringence at the locus}\label{isec5}

Photons in nonlinear electrodynamics do not follow null geodesics of \eqref{eq:metric} but of two effective optical metrics fixed by the second derivatives of the Lagrangian \citep{Novello2000,DeLorenci2000,Obukhov2002}. Since $\mathcal{L}_{FF}$ and $\mathcal{L}_{GG}$ are unequal for the quadratic theory, the two cones separate even on a purely electric background, and the separation persists over the whole dyonic family. We derive the cones from the characteristic condition, specialize them to \eqref{eq:dyonic}, and quantify the shadow doublet that follows.

Let the discontinuity of the field strength across a characteristic surface be $f_{\mu\nu}=k_{\mu}e_{\nu}-k_{\nu}e_{\mu}$. The Bianchi identity kills the term carrying $\mathcal{L}_{G}$, and the field equation leaves
\begin{equation}
\mathcal{L}_{F}k^{2}e^{\nu}
+2\mathcal{L}_{FF}\left(F^{\alpha\beta}f_{\alpha\beta}\right)k_{\mu}F^{\mu\nu}
+2\mathcal{L}_{GG}\left(\tilde{F}^{\alpha\beta}f_{\alpha\beta}\right)k_{\mu}\tilde{F}^{\mu\nu}=0 ,
\label{eq:hadamard}
\end{equation}
where the mixed derivative has been dropped because it vanishes identically by \eqref{eq:Lderivs}. Writing $X^{\nu}=k_{\mu}F^{\mu\nu}$ and $Y^{\nu}=k_{\mu}\tilde{F}^{\mu\nu}$ and contracting \eqref{eq:hadamard} with $X_{\nu}$ and with $Y_{\nu}$ gives a homogeneous pair whose determinant must vanish,
\begin{equation}
\left[\frac{\mathcal{L}_{F}k^{2}}{2}+2\mathcal{L}_{FF}X^{2}\right]
\left[\frac{\mathcal{L}_{F}k^{2}}{2}+2\mathcal{L}_{GG}Y^{2}\right]
=4\mathcal{L}_{FF}\mathcal{L}_{GG}\left(X\cdot Y\right)^{2} .
\label{eq:fresnel}
\end{equation}
In the orthonormal frame adapted to \eqref{eq:metric} the dyonic background has a radial electric field $E=F_{rt}=w^{2}$ and a radial magnetic field $B=p/r^{2}$, so $F=2(B^{2}-E^{2})$, $G=-4EB$, and the three contractions reduce to $X^{2}=E^{2}u+B^{2}v$, $Y^{2}=B^{2}u+E^{2}v$ and $X\cdot Y=\tfrac14 G\,k^{2}$, where $u=k_{\hat{t}}^{2}-k_{\hat{r}}^{2}$ and $v=k_{\hat{\theta}}^{2}+k_{\hat{\varphi}}^{2}$. Equation \eqref{eq:fresnel} then becomes a quadratic in $v/k^{2}$ whose two roots $\gamma_{\pm}$ define
\begin{equation}
\mathcal{N}_{\pm}=\left(1-\frac{1}{\gamma_{\pm}}\right)^{-1},
\qquad
\dd s_{\pm}^{2}=-f\,\dd t^{2}+\frac{\dd r^{2}}{f}+\mathcal{N}_{\pm}(r)\,r^{2}\dd\Omega^{2} ,
\label{eq:optical}
\end{equation}
the two effective metrics on whose null cones the two polarizations propagate. The quadratic has coefficients
\begin{equation}
\mathcal{A}_{2}=4\mathcal{L}_{FF}\mathcal{L}_{GG}\Sigma^{2},\quad
\mathcal{A}_{1}=2\Sigma\left(\mathcal{L}_{GG}c_{1}+\mathcal{L}_{FF}c_{2}\right),\quad
\mathcal{A}_{0}=c_{1}c_{2}-4\mathcal{L}_{FF}\mathcal{L}_{GG}E^{2}B^{2},
\label{eq:quadcoef}
\end{equation}
with $\Sigma=E^{2}+B^{2}$, $c_{1}=\tfrac12\mathcal{L}_{F}-2\mathcal{L}_{FF}E^{2}$ and $c_{2}=\tfrac12\mathcal{L}_{F}-2\mathcal{L}_{GG}B^{2}$. Both indices tend to unity as $a\to0$, where $\mathcal{L}_{FF}$ and $\mathcal{L}_{GG}$ vanish and the cones merge back onto the metric cone; we have checked this numerically to fourteen figures.

\begin{figure}[!htbp]
\centering
\includegraphics[width=0.70\textwidth]{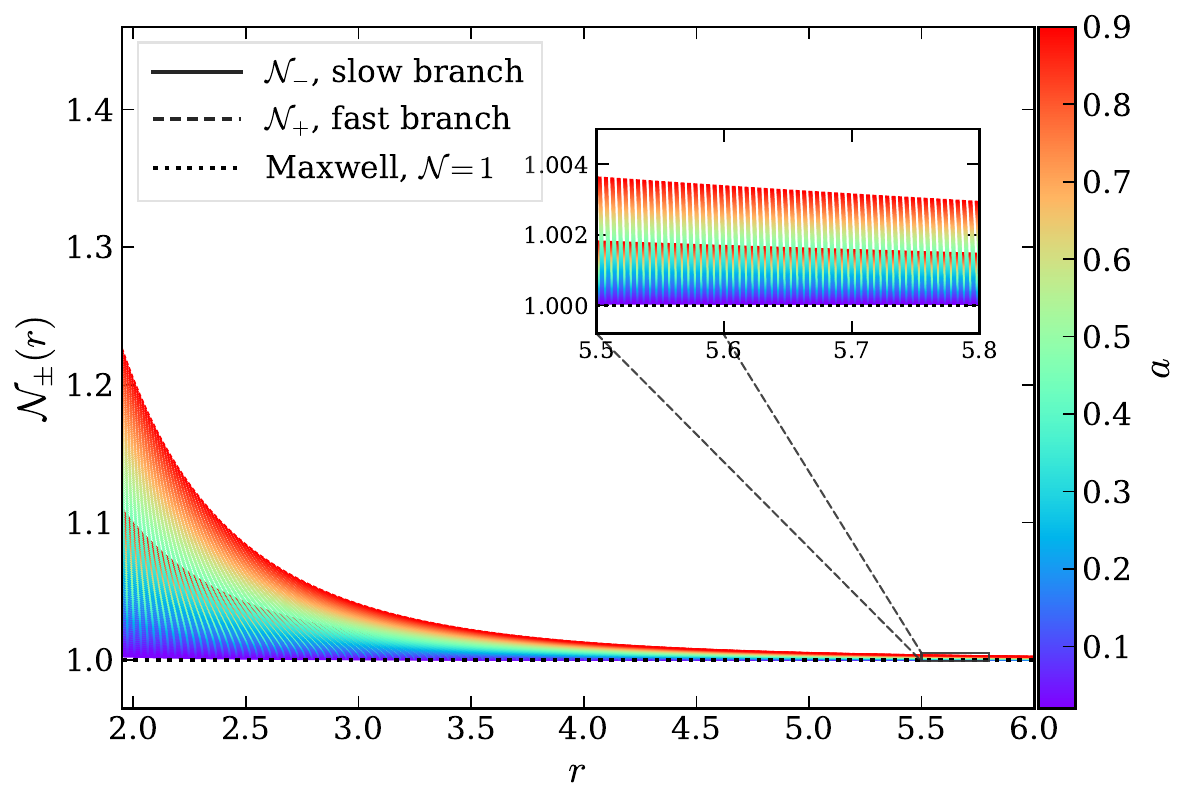}
\caption{Effective refractive indices \eqref{eq:optical} at $p=0.4$ and $q=0.55$, drawn as continuous families in the nonlinear coupling; solid curves are the slow branch and dashed curves the fast branch. Both approach unity at large radius as $r^{-4}$ and both exceed it, so the birefringence is confined to the near-horizon region. The inset resolves the two branches near the light ring.}
\label{fig:index}
\end{figure}

Because the angular part of \eqref{eq:optical} is rescaled while the radial and temporal parts are not, the photon effective potential picks up the index directly,
\begin{equation}
V_{\pm}(r)=\frac{f(r)}{\mathcal{N}_{\pm}(r)\,r^{2}} ,
\label{eq:Vphoton}
\end{equation}
so each polarization has its own light ring at $V_{\pm}'=0$ and its own critical impact parameter $b_{\pm}=V_{\pm}^{-1/2}$ evaluated there. Figure~\ref{fig:index} shows the two indices as continuous families in the coupling. Both exceed unity, so both polarizations are slower than a null geodesic of \eqref{eq:metric} would be, and the ordering $\mathcal{N}_{+}>\mathcal{N}_{-}$ holds throughout the exterior, so no polarization propagates superluminally with respect to the other cone. The indices fall off as $r^{-4}$, which is why the splitting is a near-horizon effect and not a weak-field one.

Table~\ref{tab:shadow} collects the light rings and the critical impact parameters. Three features stand out. The splitting grows roughly linearly in the coupling, reaching $0.64\%$ at $a=0.5$. It is almost independent of how the total charge is divided between the two sectors, changing by less than a fifth of its value between the purely electric and the purely magnetic configuration at fixed coupling, which is the optical counterpart of the duality symmetry of Sec.~\ref{isec3}. And it is the only place in the exterior where the coupling leaves a clear mark: the light ring of the underlying metric, which is what a massless test particle would follow and what fixes the eikonal limit of the scalar sector, barely moves at all. At $p=0.4$, $q=0.55$ and $m=1$ the geodesic critical impact parameter runs from $4.750791$ at $a\to0$ to $4.751600$ at $a=0.8$, a change of $0.017\%$, while over the same range $b_{-}$ moves to $4.818495$ and $b_{+}$ to $4.880229$, changes of $1.4\%$ and $2.7\%$. The nonlinearity is carried almost entirely by the polarization structure rather than by the geometry.

\begin{table}[!htbp]
\centering
\caption{Light rings and critical impact parameters of the two photon cones \eqref{eq:optical} at $m=1$ and $\Lambda=0$. The last column is the fractional splitting of the shadow. Configurations related by $p\leftrightarrow q$ give almost the same splitting, reflecting the duality symmetry of Sec.~\ref{isec3}.}
\label{tab:shadow}
\begin{tabularx}{\textwidth}{@{}CCCCCCCC@{}}
\toprule
{\bf $a$} & {\bf $p$} & {\bf $q$} & {\bf $r_{-}$} & {\bf $r_{+}$} & {\bf $b_{-}$} & {\bf $b_{+}$} & {\bf $100\,\Delta b/b_{-}$} \\
\midrule
$0.05$ & $0$ & $0.60$ & $2.73946$ & $2.74188$ & $4.86184$ & $4.86493$ & $0.0634$ \\
$0.05$ & $0.30$ & $0.50$ & $2.75531$ & $2.75757$ & $4.88193$ & $4.88480$ & $0.0587$ \\
$0.05$ & $0.50$ & $0.30$ & $2.75531$ & $2.75760$ & $4.88193$ & $4.88481$ & $0.0590$ \\
$0.05$ & $0.60$ & $0$ & $2.73947$ & $2.74195$ & $4.86186$ & $4.86497$ & $0.0641$ \\
$0.15$ & $0$ & $0.60$ & $2.74439$ & $2.75143$ & $4.86806$ & $4.87710$ & $0.1857$ \\
$0.15$ & $0.30$ & $0.50$ & $2.75987$ & $2.76653$ & $4.88769$ & $4.89616$ & $0.1735$ \\
$0.15$ & $0.60$ & $0$ & $2.74456$ & $2.75208$ & $4.86816$ & $4.87751$ & $0.1920$ \\
$0.30$ & $0$ & $0.60$ & $2.75154$ & $2.76496$ & $4.87717$ & $4.89468$ & $0.3591$ \\
$0.30$ & $0.30$ & $0.50$ & $2.76659$ & $2.77957$ & $4.89620$ & $4.91281$ & $0.3393$ \\
$0.30$ & $0.60$ & $0$ & $2.75218$ & $2.76746$ & $4.87757$ & $4.89625$ & $0.3829$ \\
$0.50$ & $0$ & $0.60$ & $2.76063$ & $2.78172$ & $4.88895$ & $4.91700$ & $0.5736$ \\
$0.50$ & $0.30$ & $0.50$ & $2.77531$ & $2.79625$ & $4.90734$ & $4.93430$ & $0.5495$ \\
$0.50$ & $0.60$ & $0$ & $2.76235$ & $2.78826$ & $4.89004$ & $4.92111$ & $0.6354$ \\
\bottomrule
\end{tabularx}
\end{table}

\begin{figure}[!htbp]
\centering
\includegraphics[width=0.70\textwidth]{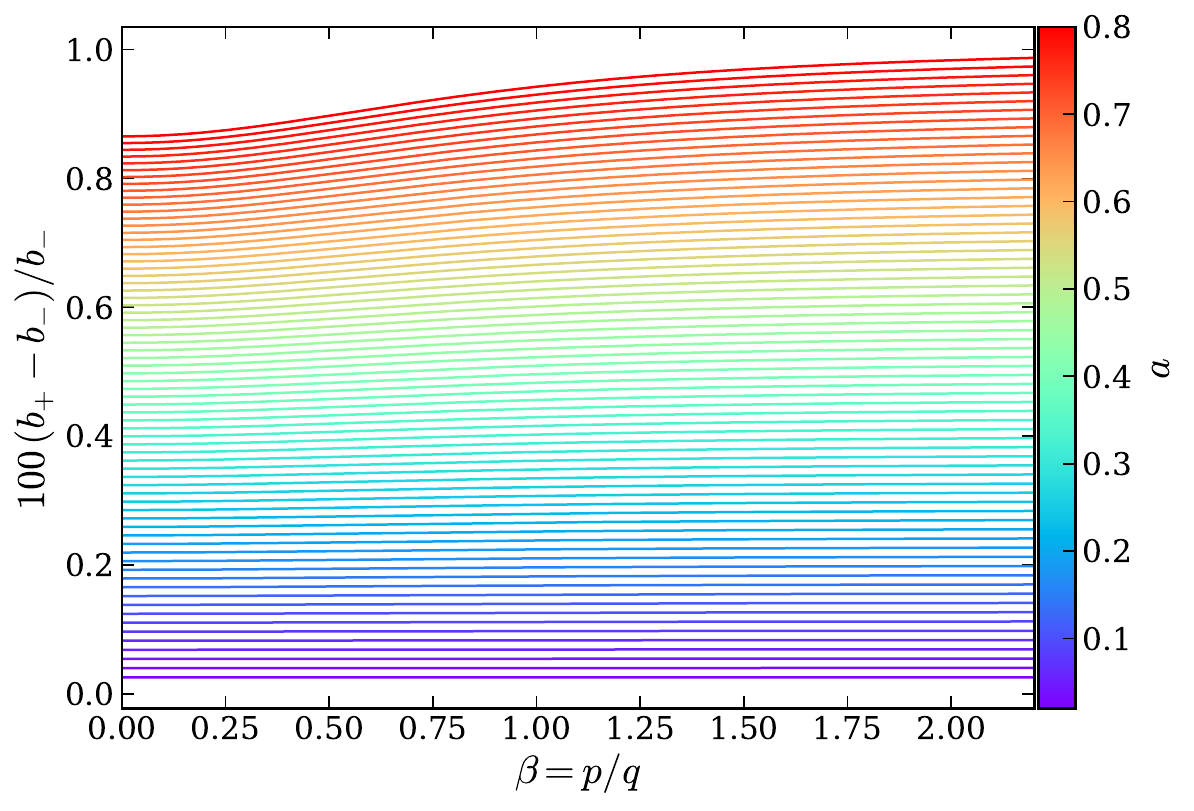}
\caption{Fractional splitting of the two critical impact parameters against the charge ratio at fixed total charge $\sqrt{p^{2}+q^{2}}=0.6$ and $m=1$, drawn as a continuous family in the nonlinear coupling. The splitting is set almost entirely by the coupling and only weakly by how the charge is divided.}
\label{fig:shadowsplit}
\end{figure}

Figure~\ref{fig:shadowsplit} traces the splitting across the charge ratio at fixed total charge. The curves are nearly flat, dipping slightly near $\beta=1$ where the two sectors contribute equally, which again mirrors the duality structure. Two shadows rather than one is a signature that has been discussed for other nonlinear theories \citep{Paula2026,GuzmanHerrera2024,GuzmanHerrera2024a,Tlemissov2025}, and the numbers here place the effect at the sub-percent level for couplings of order the horizon area, well below the roughly ten percent precision of current horizon-scale imaging but within reach of a polarization-resolved measurement that compares the two rings against each other rather than against an absolute scale.

\section{Ringdown, greybody factors and what a shadow measurement would see}\label{isec6}

\begin{figure}[!htbp]
\centering
\includegraphics[width=0.70\textwidth]{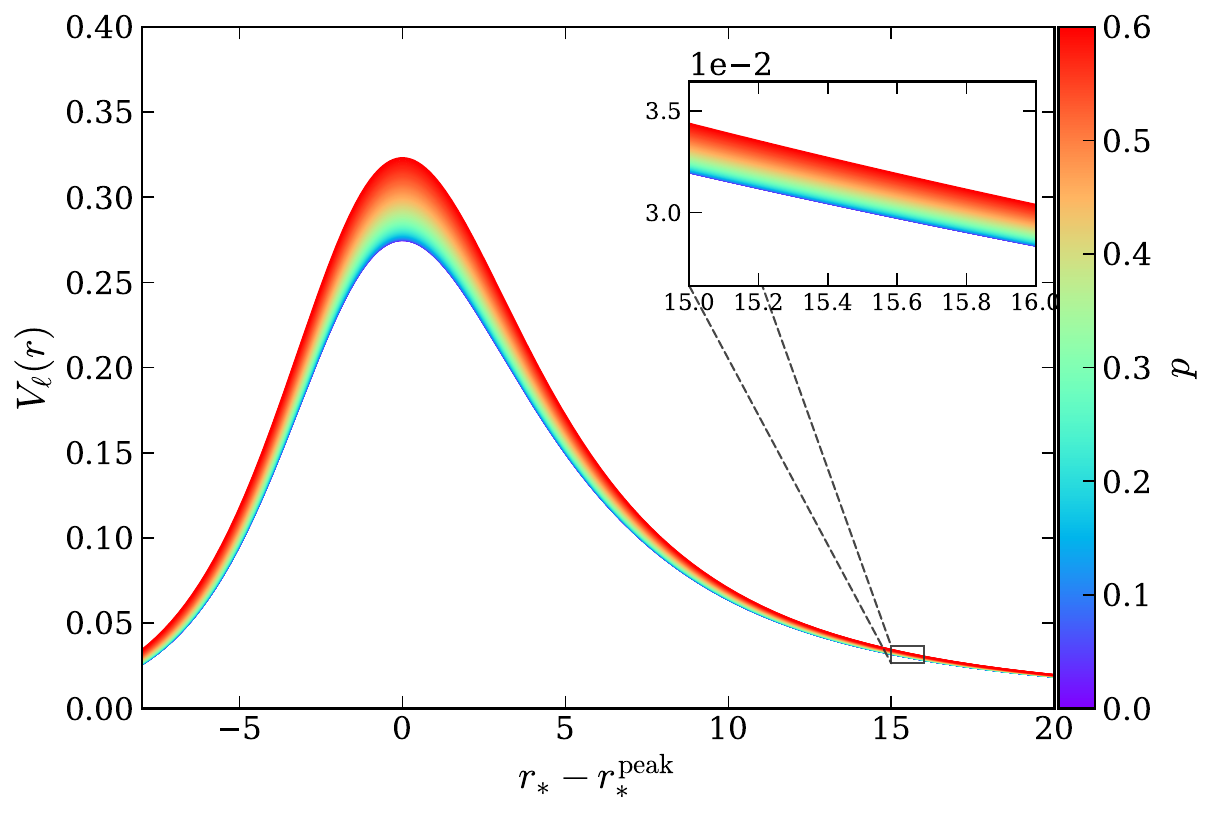}
\caption{Effective potential of \eqref{eq:radial} for $\ell=2$ at $a=0.3$, $q=0.55$, $m=1$ and $\Lambda=0$, drawn as a continuous family in the magnetic charge and aligned on the peak. The barrier rises and narrows with $p$, which is what shifts the fundamental mode.}
\label{fig:qnmpotential}
\end{figure}

The horizon phases of Sec.~\ref{isec3a} and the birefringence of Sec.~\ref{isec5} both raise the question of what an external observer could detect. We answer it for three probes: the quasinormal spectrum of a massless scalar, the transmission of that scalar through the potential barrier, and the shadow. A massless scalar field on \eqref{eq:dyonic} is minimally coupled and therefore blind to the optical metrics, so it separates in the usual way and gives
\begin{equation}
\frac{\dd^{2}R}{\dd r_{*}^{2}}+\left[\omega^{2}-V_{\ell}(r)\right]R=0,
\qquad
V_{\ell}(r)=f(r)\left[\frac{\ell(\ell+1)}{r^{2}}+\frac{f'(r)}{r}\right],
\qquad \frac{\dd r_{*}}{\dd r}=\frac{1}{f(r)} ,
\label{eq:radial}
\end{equation}
with $f$ and $f'$ supplied parametrically by \eqref{eq:dyonic} and \eqref{eq:fprime}. The barrier is shown in Fig.~\ref{fig:qnmpotential} as a continuous family in the magnetic charge, aligned on its peak so that the change in height and width is visible directly. Increasing $p$ raises the barrier and narrows it, which is the behaviour that pushes the real part of the frequency up and the damping down.

Frequencies were obtained from the third-order Wentzel--Kramers--Brillouin formula of Schutz, Will and Iyer, with the derivatives of the potential at the peak taken from a local polynomial fit in the tortoise coordinate \citep{Konoplya2011}. Against Leaver's Schwarzschild values the method returns $0.483191-0.096706i$ for the fundamental $\ell=2$ scalar mode where the continued fraction gives $0.483644-0.096759i$, an error below one part in a thousand, which sets the accuracy of everything in Table~\ref{tab:qnm}. That table carries the surprise of this section: across the whole range of couplings examined, at fixed mass and fixed charges, the fundamental $\ell=2$ frequency moves by six parts in $10^{5}$ in its real part and by three parts in $10^{3}$ in its imaginary part. The $\ell=3$ fundamental and first overtone move by comparably small amounts. The scalar ringdown is, for practical purposes, blind to the nonlinear coupling.

\begin{table}[!htbp]
\centering
\small
\caption{Scalar quasinormal frequencies of \eqref{eq:dyonic} at $p=0.40$, $q=0.55$, $m=1$ and $\Lambda=0$. The horizon radius drifts outward with the coupling by less than a part in $10^{3}$ and the frequencies drift by less than that. The method reproduces Leaver's Schwarzschild values to better than one part in $10^{3}$.}
\label{tab:qnm}
\begin{tabularx}{\textwidth}{@{}CCCCC@{}}
\toprule
{\bf $a$} & {\bf $\rH$} & {\bf $\omega_{2,0}$} & {\bf $\omega_{3,0}$} & {\bf $\omega_{3,1}$} \\
\midrule
$\to0$ & $1.733144$ & $0.528740-0.098302i$ & $0.738665-0.098264i$ & $0.725565-0.297034i$ \\
$0.10$ & $1.733494$ & $0.528764-0.098484i$ & $0.738661-0.098352i$ & $0.725747-0.297537i$ \\
$0.30$ & $1.734169$ & $0.528776-0.098657i$ & $0.738636-0.098398i$ & $0.725773-0.297726i$ \\
$0.50$ & $1.734815$ & $0.528736-0.098554i$ & $0.738591-0.098290i$ & $0.725412-0.296972i$ \\
$0.80$ & $1.735739$ & $0.528713-0.098604i$ & $0.738567-0.098459i$ & $0.725709-0.297890i$ \\
\bottomrule
\end{tabularx}
\end{table}
\normalsize

\begin{figure}[!htbp]
\centering
\includegraphics[width=0.70\textwidth]{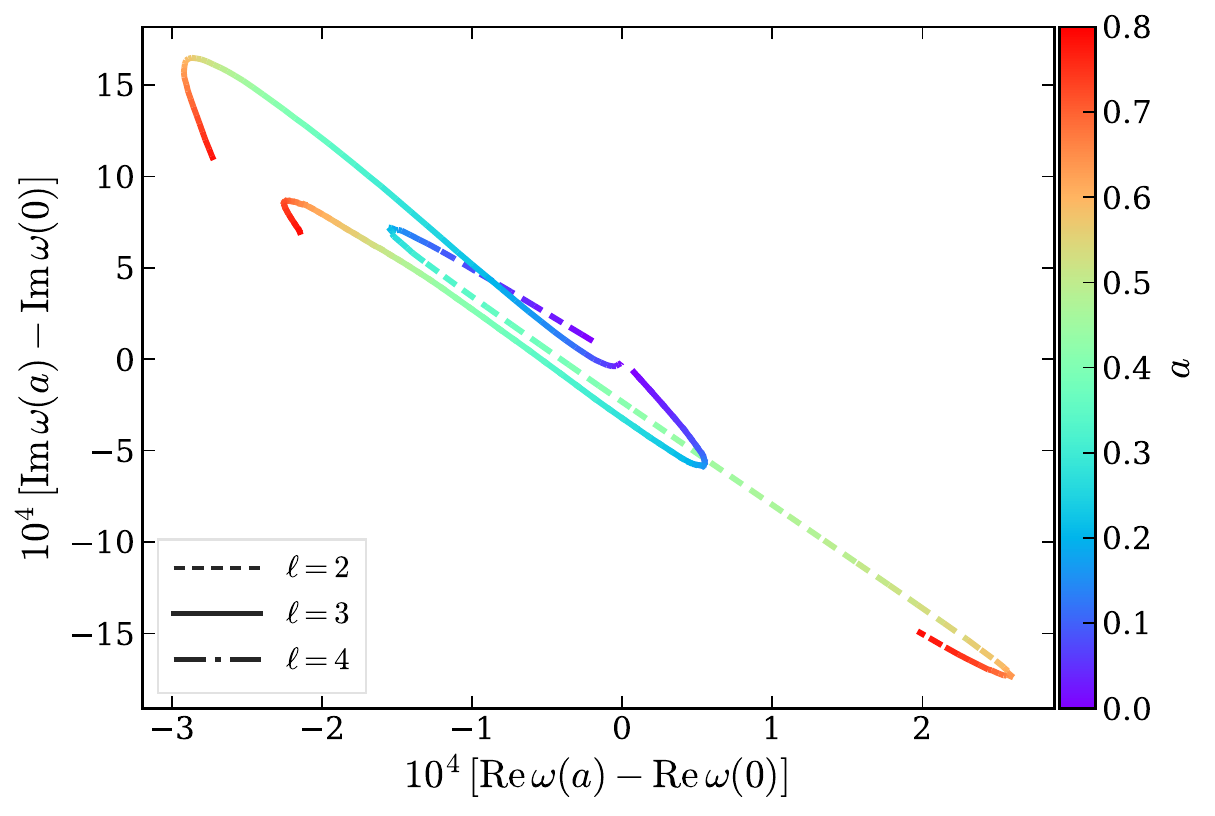}
\caption{Scalar quasinormal frequencies of \eqref{eq:dyonic} at $p=0.40$, $q=0.55$ and $m=1$, traced as the nonlinear coupling runs from zero to $0.8$; line style distinguishes the multipole. Each trajectory is shorter than the width of the symbol used to draw it would suggest, which is the statement that the minimally coupled sector is nearly blind to the coupling.}
\label{fig:qnm}
\end{figure}

The eikonal limit explains why. For a minimally coupled field the large-$\ell$ frequency is fixed by the light ring of the metric itself, $\omega\simeq\Omega_{c}(\ell+\tfrac12)-i(n+\tfrac12)|\lambda|$, and the metric light ring is what barely moves. At $p=0.40$ and $q=0.55$ the orbital frequency runs from $\Omega_{c}=0.210491$ at $a\to0$ to $0.210455$ at $a=0.8$. The Wentzel--Kramers--Brillouin frequencies converge onto that value from above as $\ell$ grows, giving $3.473548$ against $3.473106$ at $\ell=16$ and $a\to0$, an agreement of one part in $10^{4}$. Figure~\ref{fig:qnm} shows the trajectories in the complex plane as the coupling is varied, and their extent is the point: they are short. The distance travelled is comparable to the numerical uncertainty of the method, so a ringdown measurement in a minimally coupled channel cannot distinguish this theory from Einstein--Maxwell at fixed mass and charge. What it can distinguish, if the charges are known independently, is the charge combination itself, since the two enter the barrier differently.

\begin{figure}[!htbp]
\centering
\includegraphics[width=0.70\textwidth]{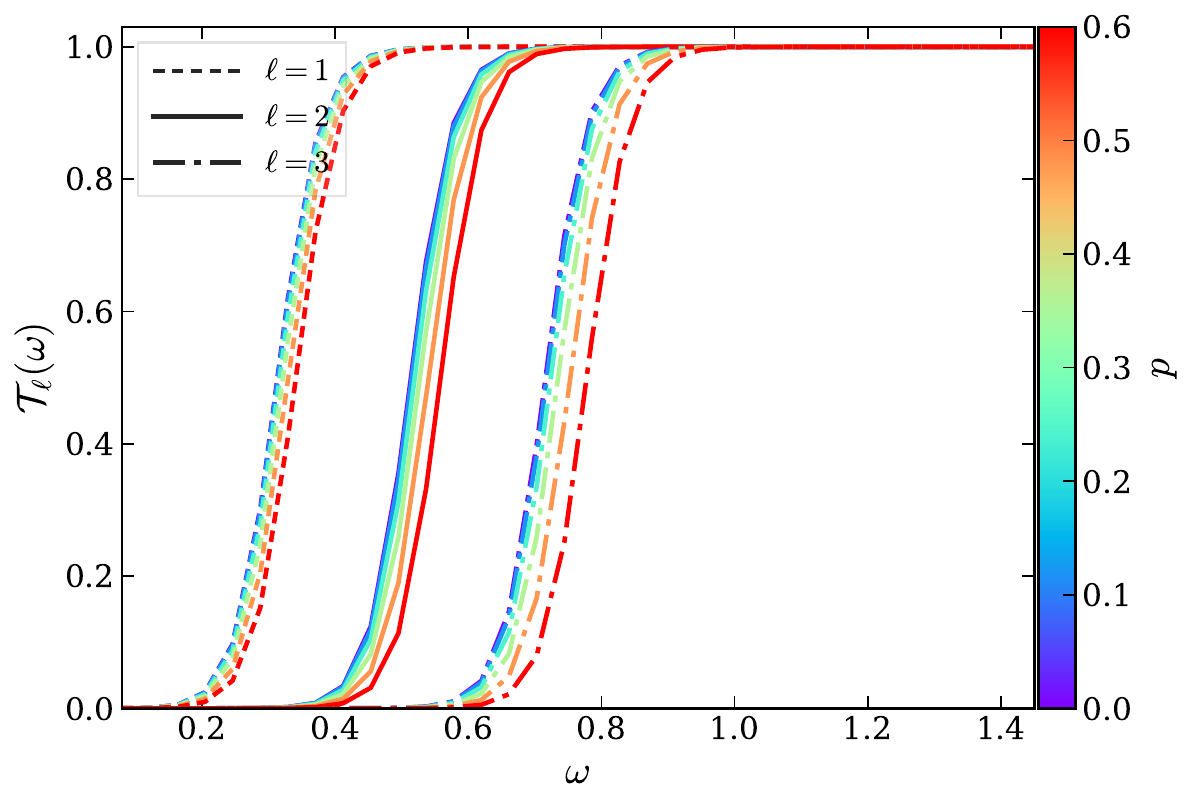}
\caption{Transmission probability for \eqref{eq:radial} at $a=0.3$, $q=0.55$ and $m=1$, drawn as a continuous family in the magnetic charge; line style distinguishes the multipole. The magnetic charge shifts the curves substantially while the nonlinear coupling, at fixed charges, does not.}
\label{fig:greybody}
\end{figure}

The same conclusion follows from the transmission through the barrier. Integrating \eqref{eq:radial} from the horizon with a purely ingoing boundary condition and reading off the incident amplitude at large radius gives the transmission probability $\mathcal{T}_{\ell}(\omega)$ plotted in Fig.~\ref{fig:greybody}. The dependence on the magnetic charge is strong, with $\mathcal{T}_{2}$ at $\omega=0.6$ falling from $0.937764$ at $p=0$ to $0.917505$ at $p=0.3$ and $0.788668$ at $p=0.6$, so the barrier that Fig.~\ref{fig:qnmpotential} shows rising has a clear effect on the emitted spectrum. The dependence on the coupling is not: at $p=0.40$, $q=0.55$ and $\omega=0.5$ the transmission is $0.266273$ at $a\to0$, $0.266554$ at $a=0.3$ and $0.267015$ at $a=0.8$. The lower bound of Visser and of Boonserm and Visser \citep{Visser1999,BoonsermVisser2008}, obtained from the integral of the potential over the tortoise coordinate, tracks the same trend but is far from saturated in the interesting range, as is usual for barriers of this shape \citep{Gray2016,BadawiGFepjc,Kanzi2020,Belchior2026,BadawiMOGepjc}.

Taken together with Sec.~\ref{isec5}, the picture is consistent and slightly unusual. Three of the four exterior observables we have examined, namely the outer horizon radius, the scalar ringdown and the greybody spectrum, are set by the metric and respond to the nonlinear coupling only through the tiny shift it induces in the lapse function at fixed mass and charge. The fourth, the shadow, is set by the optical metrics and responds directly. A single-polarization shadow measurement would inherit the insensitivity of the geometry, since $b_{-}$ and $b_{+}$ both sit within a few percent of the geodesic value and the absolute scale is degenerate with the mass. A measurement that resolves the two polarizations separately would not, because the ratio $b_{+}/b_{-}$ depends on the coupling and on nothing else at fixed charge. For the couplings of Table~\ref{tab:shadow} that ratio departs from unity by between $0.06\%$ and $0.64\%$, and it is monotone in $a$, so it functions as a clean estimator. Whether such a measurement is feasible is a separate question, and the honest answer is that it is not with present horizon-scale imaging; the relevance of the numbers is that they identify which observable carries the signal rather than that they predict a detection.

\section{Conclusion}\label{isec7}

The quadratic nonlinear electrodynamics defined by $-F+aF^{2}+bG^{2}$ has an exactly soluble dyonic sector, and it sits at $b=a/2$. On that locus the coefficient that ties the electric field equation to the magnetic charge vanishes identically, the cubic for the field strength acquires constant coefficients, and the areal radius becomes an algebraic function of the field strength. Trading one for the other turns the quadrature that fixes the lapse function into an ordinary integral of an algebraic function, which evaluates to three Gauss hypergeometric terms, and the resulting metric function \eqref{eq:dyonic} carries both charges, both couplings and the cosmological constant with no expansion anywhere. We have checked it against every limit in which the answer was independently known: dyonic Reissner--Nordstr\"om with and without a cosmological constant, the purely magnetic solution, the purely electric branch obtained by a different parametrization, Schwarzschild and Schwarzschild--anti-de Sitter, the large-radius and near-core series, and direct numerical treatment of the field equations, together with a control run off the locus where the parametric form must and does fail.

What the exact solution buys is a set of statements that a perturbative treatment could not have made. The first-order lapse function at the locus is symmetric under interchange of the electric and magnetic charges, and the exact solution shows that the symmetry survives to that order and breaks at second order with a coefficient we compute. The horizon structure contains a three-horizon window whose inner root is magnetic in origin and whose middle root is electric, so it needs both charges at once and is destroyed by an excess of either, and the outer horizon crosses the window without any visible feature. The singularity at the centre obeys three different laws: $r^{-16}$ when the magnetic charge is present, $r^{-6}$ with a shifted mass when it is not, and, at the single tuning where that shifted mass vanishes, a relegated $r^{-16/3}$ divergence weaker than Schwarzschild's. The energy density and the pressures close in elementary functions of the field variable, and the dominant energy condition reduces to one inequality in which the electric charge appears with a positive sign, so it relaxes the purely magnetic bound $\rH^{4}>4ap^{2}$ rather than tightening it. The Smarr integral for the electric branch, previously reported as having no closed form, is three hypergeometric terms in the same variable.

The thermodynamics closes with five variables. The temperature is elementary despite the hypergeometric mass, the magnetic and electric potentials are elementary and hypergeometric respectively, and the conjugate of the nonlinear coupling requires one further quadrature, itself hypergeometric, which we give in closed form. Both the first law and the Smarr formula hold numerically to the precision carried, in the full five-parameter problem and in each of its limits. In the extended phase space the coupling lowers the critical ratio monotonically below the van der Waals value of three eighths, and in magnetically dominated configurations it eventually removes the critical point altogether, at $a$ between $0.14$ and $0.16$ for a purely magnetic charge of $0.5$ and between $0.45$ and $0.50$ for the mixed configuration we examined, while the purely electric case retains its transition well beyond.

The optics separate cleanly from the geometry. Because $\mathcal{L}_{FF}$ and $\mathcal{L}_{GG}$ are unequal in the quadratic theory, the characteristic surfaces split into two cones, each with its own refractive index, its own light ring and its own critical impact parameter, and the split survives even when only one of the two charges is present. Both indices exceed unity and both fall off as the fourth power of the radius, so the effect lives near the horizon. The resulting shadow is a doublet whose fractional splitting reaches roughly half a percent for couplings comparable to the horizon area, and which barely cares how the total charge is divided between the sectors, again reflecting the near-duality of the locus. Against this, the observables that live on the metric rather than on the optical metrics are almost inert: the geodesic light ring shifts by less than two parts in $10^{4}$ over the whole coupling range we examined, the scalar quasinormal frequencies by comparable amounts, and the greybody transmission in the fourth digit. The nonlinear coupling, in other words, hides in the geometry and shows itself in the polarization structure, which is a useful thing to know when deciding what to measure.

Several tasks follow naturally and are within reach. The electromagnetic perturbations of \eqref{eq:dyonic} propagate on the two optical metrics rather than on the metric itself, so their quasinormal spectrum should inherit the birefringence that the scalar sector lacks, and computing it would turn the sub-percent shadow splitting into a ringdown observable with different systematics; the effective potentials for that calculation follow directly from \eqref{eq:optical}. The Penrose structure of the three-horizon window has not been drawn, and since the innermost root is magnetic and the middle root electric, the causal diagram and the associated mass-inflation behaviour at the inner horizon may differ from the standard charged case. The rotating counterpart is the obvious next construction; the Newman--Janis route is not available for a nonlinear source, but the slow-rotation expansion is, and the locus should survive it because the constitutive relation there is insensitive to the magnetic sector. On the thermodynamic side, the threshold at which the magnetic charge destroys the critical point deserves to be located analytically rather than numerically, and the topological classification of the resulting phase structure, which has proved informative for other dyonic anti-de Sitter families, would say whether the disappearance is a genuine change of class or a boundary effect. Finally, the same change of variable applies unchanged to the cubic that arises in any theory whose Lagrangian is quadratic in a single invariant, so the family of exactly soluble dyonic loci may be larger than the one point identified here, and mapping it is a well-posed algebraic question.

\footnotesize

\section*{Acknowledgments}

F.A. thanks the Inter University Centre for Astronomy and Astrophysics (IUCAA), Pune, India, for granting a visiting associateship.
\.{I}.S. is grateful to Eastern Mediterranean University, T\"UB\.ITAK, ANKOS and SCOAP3 for their support, and acknowledges the networking support of COST Actions CA22113, CA21106, CA23130, CA21136 and CA23115.

\section*{Data Availability Statement}

No new data were generated or analysed in support of this research. The computational worksheet used for the checks reported in Appendix~\ref{app:B} is available from the corresponding author on request.

\normalsize
\appendix

\section{Closed forms of the radial integrals and of the metric derivatives}\label{app:A}

Each power of $s$ in \eqref{eq:Jdef} integrates through
\begin{equation}
\int_{0}^{w}\frac{s^{n}\,\dd s}{\left(1+4as^{4}\right)^{5/2}}
=\frac{w^{n+1}}{n+1}\,{}_{2}F_{1}\!\left(\frac52,\frac{n+1}{4};\frac{n+5}{4};-4aw^{4}\right),
\label{eq:basic}
\end{equation}
so that the quadrature reads
\begin{equation}
J(w)=w\,{}_{2}F_{1}\!\left(\tfrac52,\tfrac14;\tfrac54;\zeta\right)
+\frac{18a}{5}w^{5}\,{}_{2}F_{1}\!\left(\tfrac52,\tfrac54;\tfrac94;\zeta\right)
+8a^{2}w^{9}\,{}_{2}F_{1}\!\left(\tfrac52,\tfrac94;\tfrac{13}{4};\zeta\right),
\qquad \zeta=-4aw^{4},
\label{eq:Jhyp}
\end{equation}
and the Smarr integral \eqref{eq:smarrclosed} reads
\begin{equation}
\int_{0}^{\wH}\frac{(1+2as^{4})(1+12as^{4})}{(1+4as^{4})^{5/2}}\dd s
=\wH\,{}_{2}F_{1}\!\left(\tfrac52,\tfrac14;\tfrac54;\zeta_{\rm H}\right)
+\frac{14a}{5}\wH^{5}\,{}_{2}F_{1}\!\left(\tfrac52,\tfrac54;\tfrac94;\zeta_{\rm H}\right)
+\frac{8a^{2}}{3}\wH^{9}\,{}_{2}F_{1}\!\left(\tfrac52,\tfrac94;\tfrac{13}{4};\zeta_{\rm H}\right).
\label{eq:smarrhyp}
\end{equation}
Differentiating the integrand of \eqref{eq:Jdef} with respect to the coupling at fixed $w$ collapses the numerator to $s^{4}(8+36as^{4}-144a^{2}s^{8})$ over $(1+4as^{4})^{7/2}$, which integrates by the same rule with $5/2$ replaced by $7/2$,
\begin{equation}
K(w,a)=\frac{8w^{5}}{5}\,{}_{2}F_{1}\!\left(\tfrac72,\tfrac54;\tfrac94;\zeta\right)
+4a\,w^{9}\,{}_{2}F_{1}\!\left(\tfrac72,\tfrac94;\tfrac{13}{4};\zeta\right)
-\frac{144a^{2}}{13}w^{13}\,{}_{2}F_{1}\!\left(\tfrac72,\tfrac{13}{4};\tfrac{17}{4};\zeta\right),
\label{eq:Khyp}
\end{equation}
which is what \eqref{eq:Acoup} needs. The integral \eqref{eq:Jdef} converges as $w\to\infty$, since its integrand falls off as $\tfrac94a^{-1/2}s^{-2}$, and the limit reduces to Beta functions,
\begin{equation}
J_{\infty}=\frac{1}{4}(4a)^{-1/4}B\!\left(\tfrac14,\tfrac94\right)
+\frac{9a}{2}(4a)^{-5/4}B\!\left(\tfrac54,\tfrac54\right)
+18a^{2}(4a)^{-9/4}B\!\left(\tfrac94,\tfrac14\right),
\label{eq:Jinf}
\end{equation}
with $B(x,y)=\Gamma(x)\Gamma(y)/\Gamma(x+y)$, scaling as $a^{-1/4}$; for $a=0.3$ this gives $J_{\infty}=4.723899431030$. Expanding \eqref{eq:dyonic} about the origin with \eqref{eq:Jinf} gives
\begin{equation}
f\simeq1+\frac{p^{2}}{r^{2}}-\frac{2ap^{4}}{5r^{6}}-\frac{2M_{\rm eff}}{r}-\frac{9q^{4/3}}{2^{5/3}a^{1/3}}r^{-2/3},
\qquad
M_{\rm eff}=m-\frac{1}{2}q^{3/2}J_{\infty},
\label{eq:core}
\end{equation}
so the electric sector contributes no Coulomb divergence, while at large $r$
\begin{equation}
f=1-\frac{2m}{r}-\frac{\Lambda r^{2}}{3}+\frac{p^{2}+q^{2}}{r^{2}}-\frac{2a\left(p^{4}+q^{4}\right)}{5r^{6}}+\frac{16a^{2}q^{6}}{9r^{10}}+O(r^{-14}),
\label{eq:largeR}
\end{equation}
whose first-order truncation is \eqref{eq:fpertlocus} and whose $r^{-10}$ term carries the duality breaking discussed in Sec.~\ref{isec3}.

Every geometric quantity in the body of the paper is built from the first two derivatives of the lapse function, which the parametrization delivers in closed form. Writing $g(w)=w^{2}(1+6aw^{4})/(1+4aw^{4})$ and using $\dd w/\dd r$ from \eqref{eq:rofw},
\begin{equation}
f'=\frac{2m}{r^{2}}-\frac{2\Lambda r}{3}-\frac{2p^{2}}{r^{3}}+\frac{12ap^{4}}{5r^{7}}
-\frac{q\,g(w)}{r}-\frac{q^{3/2}J(w)}{r^{2}} ,
\label{eq:fprime}
\end{equation}
\begin{equation}
f''=-\frac{4m}{r^{3}}-\frac{2\Lambda}{3}+\frac{6p^{2}}{r^{4}}-\frac{84ap^{4}}{5r^{8}}
+\frac{w^{4}\left(2+28aw^{4}+48a^{2}w^{8}\right)}{1+12aw^{4}}
+\frac{2q\,g(w)}{r^{2}}+\frac{2q^{3/2}J(w)}{r^{3}} .
\label{eq:fpprime}
\end{equation}
At $a=0$, where $g\to q/r^{2}$ and $J\to\sqrt{q}/r$, these collapse to the dyonic Reissner--Nordstr\"om derivatives $2m/r^{2}-2(p^{2}+q^{2})/r^{3}$ and $-4m/r^{3}+6(p^{2}+q^{2})/r^{4}$, which we have verified to fourteen figures.

\section{Numerical verification}\label{app:B}

The parametric solution was checked against direct numerical treatment of \eqref{eq:cubic} and \eqref{eq:master}, in which the cubic is solved for its real root at each radius and the quadrature performed numerically. Table~\ref{tab:check} reports the comparison at representative radii for $\Lambda=0$, the cosmological term being common to both sides. The agreement is at the level of the working precision. As a control, the same comparison performed off the locus, with $p\neq0$ and $b\neq a/2$, disagrees in the first decimal, which is the behaviour \eqref{eq:loci} predicts.

The closed-form Smarr relation of Sec.~\ref{isec4} was verified independently. At $a=0.3$, $q=0.6$ and $m=1$ the horizon lies at $\rH=1.800923$ with $T_{\rm H}=0.0393764$, the integral \eqref{eq:smarrclosed} evaluates to $0.42518137464781$ by quadrature and to the same fourteen figures through \eqref{eq:smarrhyp}, and $\tfrac12 T_{\rm H}A_{\rm H}$ plus that value returns the mass to twelve decimal places.

The perturbative lapse function \eqref{eq:fpert} was checked against both exact dyonic solutions available. At $p=0.5$, $q=0.7$, $m=1$ and $r=2.5$, the residual against the exact $a=0$ solution scales as $b^{2}$ with the coefficient $2.283\times10^{-5}$, stable across $b=10^{-2}$, $10^{-3}$ and $10^{-4}$; the residual against the exact solution \eqref{eq:dyonic} at $b=a/2$ scales as $a^{2}$ with the coefficient $2.193\times10^{-5}$, stable over the same range. Both confirm that \eqref{eq:fpert} is correct through first order and that \eqref{eq:dyonic} resums it.

The closed forms \eqref{eq:rhopt} for the energy density and the tangential pressure were checked against the same quantities computed from \eqref{eq:fprime} and \eqref{eq:fpprime} through the Einstein equations, agreeing to fifteen figures at $a=0.3$, $p=0.45$, $q=0.55$ and radii $0.8$, $1.5$ and $3.0$. The core curvature laws \eqref{eq:Kmag}, \eqref{eq:Kel} and \eqref{eq:Krel} were confirmed by evaluating $Kr^{16}$, $Kr^{6}$ and $Kr^{16/3}$ over six decades in radius and checking that each approaches the predicted constant. The duality-breaking coefficient quoted in Sec.~\ref{isec3} was obtained by evaluating $f(r;p,q)-f(r;q,p)$ at four values of the coupling spanning a factor of eight and confirming that the ratio to $a^{2}$ is stable.

\begin{table}[!htbp]
\centering
\caption{Direct numerical evaluation of \eqref{eq:master} against the parametric form \eqref{eq:dyonic}, at $\Lambda=0$.}
\label{tab:check}
\begin{tabularx}{\textwidth}{@{}LCCC@{}}
\toprule
{\bf configuration} & {\bf $r$} & {\bf $f$ numerical} & {\bf $f$ parametric} \\
\midrule
dyonic, $a=0.4$, $b=0.2$, $p=0.55$, $q=0.6$, $m=1$ & $1.898011$ & $0.1294315331816$ & $0.1294315331816$ \\
dyonic, $a=0.4$, $b=0.2$, $p=0.55$, $q=0.6$, $m=1$ & $0.601148$ & $-0.9365284314971$ & $-0.9365284314971$ \\
electric, $a=0.3$, $q=0.6$, $m=1$ & $2.500000$ & $0.2575370643242$ & $0.2575370643242$ \\
electric, $a=0.3$, $q=0.6$, $m=1$ & $0.974988$ & $-0.6854521355365$ & $-0.6854521355365$ \\
electric, $a=0.3$, $q=0.6$, $m=1$ & $0.345610$ & $-2.6104552082521$ & $-2.6104552082521$ \\
control, $a=0.4$, $b=1.0$, $p=0.55$, $q=0.6$, $m=1$ & $1.200000$ & $-0.2448982$ & $-0.2170236$ \\
\bottomrule
\end{tabularx}
\end{table}

\footnotesize
\IfFileExists{unsrtnat_linkizzet.bst}{\bibliographystyle{unsrtnat_linkizzet}}{\bibliographystyle{unsrtnat}}
\bibliography{finalref}

@article{ALBADAWI2025102076,
  author        = {Al-Badawi, Ahmad and Ahmed, Faizuddin and Sakalli, Izzet},
  title         = {Particle dynamics and thermal properties in Kalb--Ramond ModMax black holes: Theoretical predictions for observational tests of exotic physics},
  journal       = {Phys. Dark Univ.},
  volume        = {50},
  pages         = {102076},
  year          = {2025},
  issn          = {2212-6864},
  doi           = {10.1016/j.dark.2025.102076},
  url           = {https://doi.org/10.1016/j.dark.2025.102076},
}

@article{BadawiMOGepjc,
    author={Al-Badawi, Ahmad},
    title={Probing regular MOG static spherically symmetric spacetime using greybody factors and quasinormal modes},
    journal={The European Physical Journal C},
    year={2023},
    volume={83},
    pages={620},
    doi={10.1140/epjc/s10052-023-11804-4},
}

@article{BadawiGFepjc,
    author={Al-Badawi, Ahmad},
    title={Greybody factors emitted by a regular black hole in a non-minimally coupled Einstein–Yang–Mills theory},
    journal={The European Physical Journal C},
    year={2023},
    volume={83},
    pages={380},
    doi={10.1140/epjc/s10052-023-11550-7},
}

@article{Born1934,
  author  = {M. Born and L. Infeld},
  title   = {{Foundations of the new field theory}},
  journal = {Proc. R. Soc. A},
  volume  = {144},
  pages   = {425},
  year    = {1934},
  doi     = {10.1098/rspa.1934.0059}
}

@article{HeisenbergEuler1936,
  author  = {W. Heisenberg and H. Euler},
  title   = {{Consequences of Dirac theory of the positron}},
  journal = {Z. Phys.},
  volume  = {98},
  pages   = {714},
  year    = {1936},
  doi     = {10.1007/BF01343663}
}

@article{AyonBeato1998,
  author  = {E. Ayon-Beato and A. Garcia},
  title   = {{Regular black hole in general relativity coupled to nonlinear electrodynamics}},
  journal = {Phys. Rev. Lett.},
  volume  = {80},
  pages   = {5056},
  year    = {1998},
  doi     = {10.1103/PhysRevLett.80.5056}
}

@article{Bronnikov2001,
  author  = {K. A. Bronnikov},
  title   = {{Regular magnetic black holes and monopoles from nonlinear electrodynamics}},
  journal = {Phys. Rev. D},
  volume  = {63},
  pages   = {044005},
  year    = {2001},
  doi     = {10.1103/PhysRevD.63.044005}
}

@article{Croney2025,
  author  = {L. Croney and R. Gregory and C. J. Ram\'irez-Valdez},
  title   = {{Black holes in nonlinear electrodynamics}},
  journal = {JHEP},
  volume  = {10},
  pages   = {013},
  year    = {2025},
  doi     = {10.1007/JHEP10(2025)013}
}

@article{Croney2026,
  author  = {L. Croney and R. Gregory and A. Gupta and C. J. Ram\'irez-Valdez},
  title   = {{Properties of black holes in nonlinear electrodynamics}},
  journal = {Phys. Rev. D},
  volume  = {114},
  pages   = {044048},
  year    = {2026},
  doi     = {10.1103/qmjh-dzz8}
}

@article{Visser1999,
  author  = {M. Visser},
  title   = {{Some general bounds for one-dimensional scattering}},
  journal = {Phys. Rev. A},
  volume  = {59},
  pages   = {427},
  year    = {1999},
  doi     = {10.1103/PhysRevA.59.427}
}

@article{BoonsermVisser2008,
  author  = {P. Boonserm and M. Visser},
  title   = {{Bounding the Bogoliubov coefficients}},
  journal = {Ann. Phys.},
  volume  = {323},
  pages   = {2779},
  year    = {2008},
  doi     = {10.1016/j.aop.2008.02.002}
}

@article{Gray2016,
  author  = {F. Gray and S. Schuster and A. Van-Brunt and M. Visser},
  title   = {{The Hawking cascade from a black hole is extremely sparse}},
  journal = {Class. Quantum Grav.},
  volume  = {33},
  pages   = {115003},
  year    = {2016},
  doi     = {10.1088/0264-9381/33/11/115003}
}

@article{Gunasekaran2012,
  author  = {S. Gunasekaran and R. B. Mann and D. Kubiznak},
  title   = {{Extended phase space thermodynamics for charged and rotating black holes and Born-Infeld vacuum polarization}},
  journal = {JHEP},
  volume  = {11},
  pages   = {110},
  year    = {2012},
  doi     = {10.1007/JHEP11(2012)110}
}

@article{Konoplya2011,
  author  = {R. A. Konoplya and A. Zhidenko},
  title   = {{Quasinormal modes of black holes: from astrophysics to string theory}},
  journal = {Rev. Mod. Phys.},
  volume  = {83},
  pages   = {793},
  year    = {2011},
  doi     = {10.1103/RevModPhys.83.793}
}

@article{Luo2026,
  author  = {H. Luo and N. Cao and X. Y. Chew and K.-G. Lim and C. Chen and D.-h. Yeom},
  title   = {{Purely Electric, Magnetic, and Dyonic Black Holes in Einstein-Euler-Heisenberg Theory}},
  journal = {arXiv preprint},
  year    = {2026},
  eprint  = {2607.21938},
  archivePrefix = {arXiv},
  primaryClass  = {gr-qc},
  url     = {https://arxiv.org/abs/2607.21938}
}

@article{Wang2025,
  author  = {C.-H. Wang and Y.-P. Zhang and T. Zhu and S.-W. Wei},
  title   = {{Black holes in quasitopological electromagnetism}},
  journal = {arXiv preprint},
  year    = {2025},
  eprint  = {2508.20558},
  archivePrefix = {arXiv},
  primaryClass  = {gr-qc},
  url     = {https://arxiv.org/abs/2508.20558}
}

@article{Yajima2001,
  author  = {H. Yajima and T. Tamaki},
  title   = {{Black hole solutions in Euler-Heisenberg theory}},
  journal = {Phys. Rev. D},
  volume  = {63},
  pages   = {064007},
  year    = {2001},
  doi     = {10.1103/PhysRevD.63.064007}
}

@article{Magos2020,
  author  = {D. Magos and N. Bret\'on},
  title   = {{Thermodynamics of the Euler-Heisenberg-AdS black hole}},
  journal = {Phys. Rev. D},
  volume  = {102},
  pages   = {084011},
  year    = {2020},
  doi     = {10.1103/PhysRevD.102.084011}
}

@article{Breton2021,
  author  = {N. Bret\'on and L. A. L\'opez},
  title   = {{Birefringence and quasinormal modes of the Einstein-Euler-Heisenberg black hole}},
  journal = {Phys. Rev. D},
  volume  = {104},
  pages   = {024064},
  year    = {2021},
  doi     = {10.1103/PhysRevD.104.024064}
}

@article{Gulin2018,
  author  = {L. Gulin and I. Smoli\'c},
  title   = {{Generalizations of the Smarr formula for black holes with nonlinear electromagnetic fields}},
  journal = {Class. Quantum Grav.},
  volume  = {35},
  pages   = {025015},
  year    = {2018},
  doi     = {10.1088/1361-6382/aa9ee1}
}

@article{Novello2000,
  author  = {M. Novello and V. A. De Lorenci and J. M. Salim and R. Klippert},
  title   = {{Geometrical aspects of light propagation in nonlinear electrodynamics}},
  journal = {Phys. Rev. D},
  volume  = {61},
  pages   = {045001},
  year    = {2000},
  doi     = {10.1103/PhysRevD.61.045001}
}

@article{DeLorenci2000,
  author  = {V. A. De Lorenci and R. Klippert and M. Novello and J. M. Salim},
  title   = {{Light propagation in non-linear electrodynamics}},
  journal = {Phys. Lett. B},
  volume  = {482},
  pages   = {134},
  year    = {2000},
  doi     = {10.1016/S0370-2693(00)00522-0}
}

@article{Obukhov2002,
  author  = {Y. N. Obukhov and G. F. Rubilar},
  title   = {{Fresnel analysis of the wave propagation in nonlinear electrodynamics}},
  journal = {Phys. Rev. D},
  volume  = {66},
  pages   = {024042},
  year    = {2002},
  doi     = {10.1103/PhysRevD.66.024042}
}

@article{Mignemi2022,
  author  = {S. Mignemi},
  title   = {{Dyonic black holes in nonlinear electrodynamics from Kaluza-Klein theory with a Gauss-Bonnet term}},
  journal = {Int. J. Mod. Phys. A},
  volume  = {37},
  pages   = {2250065},
  year    = {2022},
  doi     = {10.1142/S0217751X22500658}
}

@article{Bronnikov2017,
  author  = {K. A. Bronnikov},
  title   = {{Dyonic Configurations in Nonlinear Electrodynamics Coupled to General Relativity}},
  journal = {Gravit. Cosmol.},
  volume  = {23},
  pages   = {343},
  year    = {2017},
  doi     = {10.1134/S0202289317040053}
}

@article{Tsuda2026,
  author  = {R. Tsuda and R. Suzuki and S. Tomizawa},
  title   = {{Existence conditions of nonsingular dyonic black holes in nonlinear electrodynamics}},
  journal = {Phys. Scr.},
  volume  = {101},
  pages   = {105005},
  year    = {2026},
  doi     = {10.1088/1402-4896/ae4b97}
}

@article{Yang2022,
  author  = {Y. Yang},
  title   = {{Dyonically charged black holes arising in generalized Born-Infeld theory of electromagnetism}},
  journal = {Ann. Phys. (N.Y.)},
  volume  = {443},
  pages   = {168996},
  year    = {2022},
  doi     = {10.1016/j.aop.2022.168996}
}

@article{Kruglov2019,
  author  = {S. I. Kruglov},
  title   = {{Dyonic Black Holes with Nonlinear Logarithmic Electrodynamics}},
  journal = {Gravit. Cosmol.},
  volume  = {25},
  pages   = {190},
  year    = {2019},
  doi     = {10.1134/S0202289319020105}
}

@article{Li2016,
  author  = {S. Li and H. Lu and H. Wei},
  title   = {{Dyonic (A)dS black holes in Einstein-Born-Infeld theory in diverse dimensions}},
  journal = {JHEP},
  pages   = {004},
  year    = {2016},
  doi     = {10.1007/JHEP07(2016)004}
}

@article{Kruglov2019a,
  author  = {I. S. Kruglov},
  title   = {{Dyonic black holes in framework of Born-Infeld-type electrodynamics}},
  journal = {Gen. Relativ. Gravit.},
  volume  = {51},
  pages   = {121},
  year    = {2019},
  doi     = {10.1007/s10714-019-2603-5}
}

@article{Ahmed2026,
  author  = {F. Ahmed and E. O. Silva},
  title   = {{Dyonic ModMax black holes in Kalb-Ramond gravity with a cloud of strings: Shadow and thermodynamics}},
  journal = {Phys. Lett. B},
  volume  = {880},
  pages   = {140769},
  year    = {2026},
  doi     = {10.1016/j.physletb.2026.140769}
}

@article{Kokoska2021,
  author  = {D. Kokoska and M. Ortaggio},
  title   = {{Static and radiating dyonic black holes coupled to conformally invariant electrodynamics in higher dimensions}},
  journal = {Phys. Rev. D},
  volume  = {104},
  pages   = {124051},
  year    = {2021},
  doi     = {10.1103/PhysRevD.104.124051}
}

@article{Acuna2026,
  author  = {G. A. Acuna and C. Bejarano and R. Ferraro},
  title   = {{Born-Infeld electrogravity and dyonic black holes}},
  journal = {Class. Quantum Grav.},
  volume  = {43},
  pages   = {155020},
  year    = {2026},
  doi     = {10.1088/1361-6382/ae9114}
}

@article{Yang2023,
  author  = {Y. Yang},
  title   = {{Dyonic matter equations, exact point-source solutions, and charged black holes in generalized Born-Infeld theory}},
  journal = {Phys. Rev. D},
  volume  = {107},
  pages   = {085007},
  year    = {2023},
  doi     = {10.1103/PhysRevD.107.085007}
}

@article{Bokuli2026,
  author  = {A. Bokuli and F. Po},
  title   = {{Noncommutative dyonic black holes sourced by nonlinear electromagnetic fields}},
  journal = {Phys. Rev. D},
  volume  = {113},
  pages   = {084026},
  year    = {2026},
  doi     = {10.1103/t8k4-1thf}
}

@article{Bronnikov2018,
  author  = {K. A. Bronnikov},
  title   = {{Nonlinear electrodynamics, regular black holes and wormholes}},
  journal = {Int. J. Mod. Phys. D},
  volume  = {27},
  pages   = {1841005},
  year    = {2018},
  doi     = {10.1142/S0218271818410055}
}

@article{Amaro2020,
  author  = {D. Amaro and A. Macias},
  title   = {{Geodesic structure of the Euler-Heisenberg static black hole}},
  journal = {Phys. Rev. D},
  volume  = {102},
  pages   = {104054},
  year    = {2020},
  doi     = {10.1103/PhysRevD.102.104054}
}

@article{Amaro2022,
  author  = {D. Amaro and A. Macias},
  title   = {{Exact lens equation for the Einstein-Euler-Heisenberg static black hole}},
  journal = {Phys. Rev. D},
  volume  = {106},
  pages   = {064010},
  year    = {2022},
  doi     = {10.1103/PhysRevD.106.064010}
}

@article{Breton2022,
  author  = {N. Breton and C. Laemmerzahl and A. Macias},
  title   = {{Rotating structure of the Euler-Heisenberg black hole}},
  journal = {Phys. Rev. D},
  volume  = {105},
  pages   = {104046},
  year    = {2022},
  doi     = {10.1103/PhysRevD.105.104046}
}

@article{Nashed2021,
  author  = {G. G. L. Nashed and S. Nojiri},
  title   = {{Mimetic Euler-Heisenberg theory, charged solutions, and multihorizon black holes}},
  journal = {Phys. Rev. D},
  volume  = {104},
  pages   = {044043},
  year    = {2021},
  doi     = {10.1103/PhysRevD.104.044043}
}

@article{Ali2024,
  author  = {A. Ali and A. Ovgun},
  title   = {{Topological dyonic black holes of massive gravity with generalized quasitopological electromagnetism}},
  journal = {Eur. Phys. J. C},
  volume  = {84},
  pages   = {378},
  year    = {2024},
  doi     = {10.1140/epjc/s10052-024-12710-z}
}

@article{Chen2024,
  author  = {Z. Chen and S. Wei},
  title   = {{Thermodynamical topology with multiple defect curves for dyonic AdS black holes}},
  journal = {Eur. Phys. J. C},
  volume  = {84},
  pages   = {1294},
  year    = {2024},
  doi     = {10.1140/epjc/s10052-024-13620-w}
}

@article{Li2022a,
  author  = {M. Li and H. Wang and S. Wei},
  title   = {{Triple points and novel phase transitions of dyonic AdS black holes with quasitopological electromagnetism}},
  journal = {Phys. Rev. D},
  volume  = {105},
  pages   = {104048},
  year    = {2022},
  doi     = {10.1103/PhysRevD.105.104048}
}

@article{Sekhmani2023,
  author  = {Y. Sekhmani and D. J. Gogoi},
  title   = {{Electromagnetic quasinormal modes of dyonic AdS black holes with quasitopological electromagnetism in a Horndeski gravity theory mimicking EGB gravity at D? 4}},
  journal = {Int. J. Geom. Methods Mod. Phys.},
  volume  = {20},
  pages   = {2350160},
  year    = {2023},
  doi     = {10.1142/S0219887823501608}
}

@article{Breton2005,
  author  = {N. Bret\'on},
  title   = {{Smarr's formula for black holes with non-linear electrodynamics}},
  journal = {Gen. Relativ. Gravit.},
  volume  = {37},
  pages   = {643},
  year    = {2005},
  doi     = {10.1007/s10714-005-0051-x}
}

@article{Zhang2018,
  author  = {Y. Zhang and S. Gao},
  title   = {{First law and Smarr formula of black hole mechanics in nonlinear gauge theories}},
  journal = {Class. Quantum Grav.},
  volume  = {35},
  pages   = {145007},
  year    = {2018},
  doi     = {10.1088/1361-6382/aac9d4}
}

@article{Balart2017,
  author  = {L. Balart and S. Fernando},
  title   = {{A Smarr formula for charged black holes in nonlinear electrodynamics}},
  journal = {Mod. Phys. Lett. A},
  volume  = {32},
  pages   = {1750219},
  year    = {2017},
  doi     = {10.1142/S0217732317502194}
}

@article{Pereira2014,
  author  = {J. P. Pereira and H. J. Mosquera Cuesta and J. A. Rueda and R. Ruffini},
  title   = {{On the black hole mass decomposition in nonlinear electrodynamics}},
  journal = {Phys. Lett. B},
  volume  = {734},
  pages   = {396},
  year    = {2014},
  doi     = {10.1016/j.physletb.2014.04.047}
}

@article{Fan2016,
  author  = {Z. Fan and X. Wang},
  title   = {{Construction of regular black holes in general relativity}},
  journal = {Phys. Rev. D},
  volume  = {94},
  pages   = {124027},
  year    = {2016},
  doi     = {10.1103/PhysRevD.94.124027}
}

@article{Kruglov2022,
  author  = {S. I. Kruglov},
  title   = {{Nonlinearly charged AdS black holes, extended phase space thermodynamics and Joule-Thomson expansion}},
  journal = {Ann. Phys. (N.Y.)},
  volume  = {441},
  pages   = {168894},
  year    = {2022},
  doi     = {10.1016/j.aop.2022.168894}
}

@article{Kruglov2022a,
  author  = {S. I. Kruglov},
  title   = {{NED-AdS black holes, extended phase space thermodynamics and Joule-Thomson expansion}},
  journal = {Nucl. Phys. B},
  volume  = {984},
  pages   = {115949},
  year    = {2022},
  doi     = {10.1016/j.nuclphysb.2022.115949}
}

@article{Kuang2018,
  author  = {X. Kuang and B. Liu and A. Ovguen},
  title   = {{Nonlinear electrodynamics AdS black hole and related phenomena in the extended thermodynamics}},
  journal = {Eur. Phys. J. C},
  volume  = {78},
  pages   = {840},
  year    = {2018},
  doi     = {10.1140/epjc/s10052-018-6320-0}
}

@article{Wu2018,
  author  = {C. Wu and D. Zou and Y. Wang},
  title   = {{P-V Criticality of Born-Infeld AdS Black Holes Surrounded by Quintessence}},
  journal = {Commun. Theor. Phys.},
  volume  = {70},
  pages   = {459},
  year    = {2018},
  doi     = {10.1088/0253-6102/70/4/459}
}

@article{Gao2021,
  author  = {C. Gao},
  title   = {{Black holes with many horizons in the theories of nonlinear electrodynamics}},
  journal = {Phys. Rev. D},
  volume  = {104},
  pages   = {064038},
  year    = {2021},
  doi     = {10.1103/PhysRevD.104.064038}
}

@article{Paula2026,
  author  = {M. A. A. de Paula and H. C. D. Lima and P. V. P. Cunha and C. A. R. Herdeiro and L. C. B. Crispino},
  title   = {{Two shadows of a single black hole: Vacuum birefringence phenomena within Einstein-nonlinear-electrodynamics}},
  journal = {Phys. Rev. D},
  volume  = {114},
  pages   = {024043},
  year    = {2026},
  doi     = {10.1103/8lkf-mgjw}
}

@article{GuzmanHerrera2024,
  author  = {E. Guzman-Herrera and N. Breton},
  title   = {{Light propagation in the vicinity of the ModMax black hole}},
  journal = {JCAP},
  pages   = {041},
  year    = {2024},
  doi     = {10.1088/1475-7516/2024/01/041}
}

@article{GuzmanHerrera2024a,
  author  = {E. Guzman-Herrera and A. Montiel and N. Breton},
  title   = {{Comparative of light propagation in Born-Infeld, Euler-Heisenberg and ModMax nonlinear electrodynamics}},
  journal = {JCAP},
  pages   = {002},
  volume  = {11},
  year    = {2024},
  doi     = {10.1088/1475-7516/2024/11/002}
}

@article{Tlemissov2025,
  author  = {A. Tlemissov and B. Toshmatov and J. Kovar},
  title   = {{Effect of nonlinear electrodynamics on polarization distribution around black holes}},
  journal = {Phys. Rev. D},
  volume  = {111},
  pages   = {064084},
  year    = {2025},
  doi     = {10.1103/PhysRevD.111.064084}
}

@article{Ahmed2026a,
  author  = {F. Ahmed and A. Al-Badawi and I. Sakalli and F. Javed and S. Kanzi},
  title   = {{Imprints of ModMax electrodynamics on particle dynamics and thermodynamic properties of Kalb-Ramond AdS black holes}},
  journal = {Int. J. Geom. Methods Mod. Phys.},
  year    = {2026},
  doi     = {10.1142/S0219887826501306}
}

@article{Kanzi2020,
  author  = {S. Kanzi and S. H. Mazharimousavi and I. Sakalli},
  title   = {{Greybody factors of black holes in dRGT massive gravity coupled with nonlinear electrodynamics}},
  journal = {Ann. Phys. (N.Y.)},
  volume  = {422},
  pages   = {168301},
  year    = {2020},
  doi     = {10.1016/j.aop.2020.168301}
}

@article{Belchior2026,
  author  = {F. M. Belchior and A. R. P. Moreira and A. Bouzenada and F. Ahmed},
  title   = {{Absorption, greybody factors of bosonic particles and thermodynamic properties in ModMax-black holes sourced by global monopole in Kalb-Ramond gravity}},
  journal = {Ann. Phys. (N.Y.)},
  volume  = {492},
  pages   = {170548},
  year    = {2026},
  doi     = {10.1016/j.aop.2026.170548}
}

@article{Ortaggio2026,
  author  = {M. Ortaggio},
  title   = {{Einstein-Maxwell fields as solutions of Einstein gravity coupled to conformally invariant nonlinear electrodynamics}},
  journal = {Phys. Rev. D},
  volume  = {113},
  pages   = {L081503},
  year    = {2026},
  doi     = {10.1103/nwcq-n4hg}
}

@article{Mou2024,
  author  = {P. Mou and Q. Jiang and K. He and G. Li},
  title   = {{Triple points and phase transitions of D-dimensional dyonic AdS black holes with quasitopological electromagnetism in Einstein-Gauss-Bonnet gravity}},
  journal = {Chin. Phys. B},
  volume  = {33},
  pages   = {060401},
  year    = {2024},
  doi     = {10.1088/1674-1056/ad3342}
}

@article{Barrientos2022,
  author  = {J. Barrientos and J. Mena},
  title   = {{Joule-Thomson expansion of AdS black holes in quasitopological electromagnetism}},
  journal = {Phys. Rev. D},
  volume  = {106},
  pages   = {044064},
  year    = {2022},
  doi     = {10.1103/PhysRevD.106.044064}
}

\end{document}